\documentclass[pdftex,twocolumn,epjc3]{svjour3}          

\RequirePackage[T1]{fontenc}
\usepackage[T1]{fontenc}
\usepackage[utf8]{inputenc}
\usepackage{amsmath, amssymb}
\smartqed  
\usepackage{cuted} 
\RequirePackage{graphicx}
\RequirePackage{mathptmx}      
\RequirePackage{flushend}
\usepackage{amssymb}
\usepackage{amsmath}
\usepackage{amsfonts}
\usepackage[utf8]{inputenc}
\usepackage{cuted}
\usepackage{multicol} 
\makeatletter
\usepackage{tikz}
\usepackage{float}
\usepackage{ctable}
\definecolor{green}{rgb}{0.0, 0.56, 0.0}
\usepackage{adjustbox}
\providecommand{\openone}{\leavevmode\hbox{\small1\kern-3.8pt\normalsize1}}

\newcommand{\tb}{$(TB)$}
\newcommand{\tby}{$(TBY)$}
\def\half{\tfrac{1}{2}}
\definecolor{green}{rgb}{0.0, 0.56, 0.0}

\usepackage{etoolbox}
\usepackage{caption}

\providecommand{\openone}{\leavevmode\hbox{\small1\kern-3.8pt\normalsize1}}

\newcommand{\xtb}{$(XTB)$}
\def\half{\tfrac{1}{2}}

\makeatother
\RequirePackage[numbers,sort&compress]{natbib}
\RequirePackage[colorlinks,citecolor=blue,urlcolor=blue,linkcolor=blue]{hyperref}

\journalname{Eur. Phys. J. C}

\begin{document}

\title{Anatomy of Vector-Like Top-Quark Models in the 
	Alignment Limit \\\vspace{0.2cm} of the 2-Higgs Doublet Model Type-II}

\author{A. Arhrib\thanksref{e1,addr1,addr2}
        \and
        R. Benbrik\thanksref{e2,addr3} 
                \and
        	M. Boukidi\thanksref{e3,addr3} 
                \and
        B.~Manaut\thanksref{e4,addr4} 
                \and
        S. Moretti\thanksref{e5,addr5,addr6} 
}

\thankstext{e1}{e-mail: \href{mailto:aarhrib@gmail.com}{aarhrib@gmail.com}}
\thankstext{e2}{e-mail: \href{mailto:r.benbrik@uca.ac.ma}{r.benbrik@uca.ac.ma}}
\thankstext{e3}{e-mail: \href{mailto:mohammed.boukidi@ced.uca.ma}{mohammed.boukidi@ced.uca.ma}}
\thankstext{e4}{e-mail: \href{mailto:b.manaut@usms.ma}{b.manaut@usms.ma}}
\thankstext{e5}{e-mail: \href{mailto:stefano.moretti@physics.uu.se}{stefano.moretti@physics.uu.se}; \href{mailto:s.moretti@soton.ac.uk}{s.moretti@soton.ac.uk} }

\institute{Abdelmalek Essaadi University, Faculty of Sciences and Techniques, Tangier, Morocco\label{addr1}
          \and
          		Department of Physics and CTC, National Tsing Hua University, Hsinchu, Taiwan 300\label{addr2} 
          		\and
          Polydisciplinary Faculty, Laboratory of Fundamental and Applied Physics, Cadi Ayyad University, Sidi Bouzid, B.P. 4162, Safi, Morocco\label{addr3}
          \and
           Polydisciplinary Faculty, Laboratory of Research in Physics and Engineering Sciences,  Sultan Moulay Slimane University, Beni Mellal 23000, Morocco\label{addr4}
          \and
            Department of Physics \& Astronomy, Uppsala University, Box 516, SE-751 20 Uppsala, Sweden\label{addr5}
          \and
            School of Physics \& Astronomy, University of Southampton, Southampton, SO17 1BJ, United Kingdom\label{addr6}
}

\date{Received: date / Accepted: date}

\maketitle

\begin{abstract}
	A comprehensive extension of the ordinary 2-Higgs Doublet Model (2HDM), supplemented by Vector-Like Quarks (VLQs), in the ``alignment limit'' is presented. In such a scenario, we study the possibility that Large Hadron Collider (LHC) searches for VLQs can profile their nature too, i.e., whether they belong to a singlet, doublet, or triplet representation. To achieve this, we exploit both Standard Model (SM) decays of VLQs with top-(anti)quark Electromagnetic (EM) charge ($T$), i.e., into $b,t$ quarks and $W^\pm, Z,h$ bosons (which turn out to be suppressed and hence $T$ states can escape existing limits) as well as their exotic decays, i.e., into $b,t$ (and possibly $B$) quarks and $H^\pm, H, A$ bosons. We show that quite specific decay patterns emerge in the different VLQ representations so that, depending upon which $T$ signals are accessed at the LHC, one may be able to ascertain the underlying Beyond Standard Model (BSM) structure, especially if mass knowledge of the new fermionic and bosonic sectors can be inferred from (other) data.
\end{abstract}

	\section{Introduction} 
Following the discovery of a Higgs boson, $h$, during Run 1 of the Large Hadron Collider (LHC) at CERN \cite{ATLAS:2012yve, CMS:2012qbp}, the ATLAS and CMS collaborations have carried out a broad programme of measurements of its properties, namely, mass, spin, CP quantum numbers, etc. These all seem to point to a Standard Model (SM) nature of such a new state. However, some anomalies exist in current data that may point to a Beyond the SM (BSM) framework accommodating such a discovery. The signal strength of the $t\bar{t}h$ \cite{CMS:2022dwd,ATLAS:2022vkf} associated production mode is one of those most prominent, while milder effects are still seen in the fits to data when assuming the gluon-gluon fusion production mode, especially when combined with di-photon decays. Further, current measurements of the $h\to \gamma Z$ decay channel \cite{ATLAS:2023yqk} do not exclude that its rate can differ from  SM predictions. A possibility to capture at once all such anomalies is offered by the presence of Vector-Like Quarks (VLQs) as they could affect simultaneously the one-loop induced (chiefly by $t$ quarks) SM-like Higgs production and decay channels as well as mediate a  $t\bar t h$ final state.   

At the same time, having established the doublet nature of the SM-like Higgs field so far discovered, much experimental attention has lately been aroused by the possibility that other Higgs doublets could exist in nature. The simplest option is the one offered by a 2-Higgs Doublet Model (2HDM) \cite{Branco:2011iw}. Herein, four additional Higgs bosons exist: a heavier CP-even state, $H$, a CP-odd one, $A$, and a pair of charged ones, $H^\pm$. All such new Higgs states could then manifest themselves at the LHC in new direct signals, i.e., when they are produced as resonant objects inside the detectors, or else indirectly, e.g., via loop effects in a variety of single and double $h$ production modes. In fact, in the first case, they may well be produced in the decays of new particles.

Therefore, it is intriguing to study models that embed  2HDM (pseudo)scalar states as well as VLQs. Indeed,we concentrate here on both these states at once. While the nature of the former has already been described, it is worth reminding the reader that the latter are heavy spin 1/2 particles that transform as triplets under colour but, unlike SM quarks, their left- and right-handed couplings have the same Electroweak (EW) quantum numbers. Furthermore, their couplings to Higgs bosons do not participate in the standard EW Symmetry Breaking (EWSB)dynamics onset by the Higgs mechanism, hence, they are not of Yukawa type (i.e., proportional to the mass), rather they are additional parameters, which can then be set as needed in order to achieve both compliance with present data and predict testable signals for the future.

Amongst the ensuing signatures, an exciting possibility is constituted by the decays of VLQs into the additional Higgs states of a 2HDM, which are not currently being pursued at the LHC. (Recently, we have advocated that $\gamma\gamma$ and $Z\gamma$ signatures of a heavy neutral Higgs state produced from a heavy VLQ top state might be accessible during Run 3 \cite{Benbrik:2019zdp}.) The ATLAS and CMS collaborations, while collecting data at 7, 8, and 13 TeV, have performed searches for VLQs with different quantum numbers, probing single and pair production mechanisms but limited to decay modes into SM quarks and gauge/Higgs bosons (for the most updated experimental results of ATLAS and CMS, we refer to the respective web pages \cite{twikiATLAS8TeV,twikiATLAS13TeV,twikiCMS}). However, no evidence for the existence of other quarks, besides those of the SM, has been obtained. The reason may indeed be that additional fermions may preferentially decay into additional Higgs states, rather than via SM objects.        

Another motivation for pursuing the phenomenological exploitation of a 2HDM plus VLQs (henceforth, `2HDM+VLQ') construct finally comes from the expectation that, by leveraging the cancellations that may occur at the loop level between the bosonic (onset by the additional Higgs states) and fermionic (onset by the additional VLQ states) contributions, as well as the altered coupling structure of the SM states (chiefly, of the SM-like Higgs and third generation quarks), a larger expanse of parameter space of this model will be compatible with  EW Precision Observables (EWPOs), from LEP and SLC, than what would be released by only accounting for separate (as opposed to simultaneous) effects from, on the one hand, 2HDM Higgses or, on the other hand, VLQs. Indeed, it would be rewarding to verify that this can happen for rather light masses of both additional Higgs and VLQ states so that they can both be searched for at the LHC shortly through the aforementioned exotic decays.        

In the present paper, we wish to build on the results of \cite{Benbrik:2015fyz,Arhrib:2016rlj,Aguilar-Saavedra:2013qpa,Badziak:2015zez,Angelescu:2015uiz,Aguilar-Saavedra:2009xmz,DeSimone:2012fs,Kanemura:2015mxa,Lavoura:1992np,Chen:2017hak,Carvalho:2018jkq,Moretti:2016gkr,Prager:2017hnt,Prager:2017owg,Moretti:2017qby,Deandrea:2017rqp,Aguilar-Saavedra:2017giu,Alves:2023ufm,Dermisek:2019vkc,Dermisek:2020gbr,Dermisek:2021zjd}, but especially \cite{Benbrik:2019zdp},by studying a 2HDM plus VLQ scenario where the VLQs can belong to a singlet, doublet, or triplet representation under the SM gauge group $SU(3)_C\times SU(2)_L\times U(Y)_Y$. We intend to tension the scope afforded, in the quest for such VLQs, by their exotic decays (involving the additional Higgs states of the 2HDM, both neutral and charged, as well as other, lighter VLQs) against the one already established through their SM decays, involving standard quarks and gauge/Higgs bosons as final state products. Particularly, we will pursue the task of uniquely attributing certain VLQ decay patterns into exotic states that may be established at the LHC to one or another of the three aforementioned multiplet structures.        

Our paper is organised as follows. In the next section, we describe in detail the three theoretical structures concerned, both their model construction and implementation. In Sect. III, using three    subsections, one for each multiplet realization, we present our results.All this is followed by our conclusions, in Sect. IV. We also have an Appendix with relevant Feynman rules involving Higgs states and the new VLQs.

\section{Model description}

\subsection{Formalism}
In this paper, we extend Ref. \cite{Arhrib:2016rlj}, wherein a CP-conserving 2HDM with a singlet VLQ companion to the (chiral) top quark of the SM was set up, that contained the canonical Higgs states: as mentioned,  
two CP-even states denoted by $h$ (the lightest) and $H$ (the heaviest), one CP-odd state denoted by $A$ and two charged states denoted by $H^\pm$.
As tree-level Flavour Changing Neutral Currents (FCNCs) are very constrained by experiment, we imposed a $\mathbb{Z}_2$ symmetry, $\Phi_1 \to \Phi_1$ and $\Phi_2 \to - \Phi_2$, on the Higgs fields. The resulting Higgs potential (softly broken by the dimension two terms $\propto m^2_{12}$) can be written as 
\begin{eqnarray} \label{pot}
\mathcal{V} &=& m^2_{11}\Phi_1^\dagger\Phi_1+m^2_{22}\Phi_2^\dagger\Phi_2
-\left(m^2_{12}\Phi_1^\dagger\Phi_2+{\rm h.c.}\right)
\nonumber \\
&&+\half\lambda_1\left(\Phi_1^\dagger\Phi_1\right)^2
+\half\lambda_2\left(\Phi_2^\dagger\Phi_2\right)^2  +\lambda_3\Phi_1^\dagger\Phi_1\Phi_2^\dagger\Phi_2
\nonumber \\
&&+\lambda_4\Phi_1^\dagger\Phi_2\Phi_2^\dagger\Phi_1
+\left[\half\lambda_5\left(\Phi_1^\dagger\Phi_2\right)^2+{\rm h.c.}\right].
\end{eqnarray}
choosing real Vacuum Expectation Values (VEVs) for the two Higgs doublet fields, $v_1$ and $v_2$, and demanding
$m_{12}^2$ and $\lambda_5$ to be real, the potential is indeed CP-conserving.
The free independent parameters are here taken to be the four masses, $m_h$, $m_H$, $m_A$ and $m_{H^\pm}$,
the soft breaking parameter $m_{12}$, the VEV ratio $\tan \beta = v_2/v_1$ and the mixing term $\sin(\beta-\alpha)$,
where the angle $\alpha$ diagonalizes the CP-even mass matrix. When we impose that no (significant) tree-level FCNCs are present in the theory using the (softly broken) $\mathbb{Z}_2$ symmetry, we end up with four different Yukawa versions of the model.  These are Type-I, where only $\Phi_2$ couples to all fermions; Type-II, where $\Phi_2$ couples to up-type quarks and $\Phi_1$ couples to 
charged leptons and down-type quarks; Type-Y (or Flipped), where $\Phi_2$ couples to charged leptons and up-type quarks and $\Phi_1$ couples to down-type quarks; Type-X (or Lepton Specific), where  $\Phi_2$ couples to quarks and $\Phi_1$ couples to charged leptons\footnote{In this paper, we will be discussing only Type-II.}.

The gauge invariant structures that have multiplets with definite ${SU}(3)_C \times {SU}(2)_L \times {U}(1)_Y$ quantum numbers appear in the interactions of new VLQs with the SM states via renormalizable couplings. The set of VLQ representations is indicated by:
\begin{align}
& T_{L,R}^0 \, && \text{(singlets)} \,, \notag \\
& (X\,T^0)_{L,R} \,, \quad (T^0\,B^0)_{L,R} \, && \text{(doublets)} \,, \notag \\
& (X\,T^0\,B^0)_{L,R} \,, \quad (T^0\,B^0\,Y)_{L,R}  && \text{(triplets)} \,.
\end{align}
We use in this section a zero superscript to distinguish the weak eigenstates from the mass eigenstates. The electric charges of the new VLQs are $ Q_T = 2/3 $, $ Q_B = -1/3 $, $ Q_X = 5/3 $ and $ Q_Y = -4/3 $. Note that $T$ and $B$ carry the same electric charge as the SM top and bottom quarks, respectively. 

The physical up-type quark mass eigenstates may, in general, contain non-zero $Q_{L,R}^0$ (with $Q$ being the VLQ field) components, when new fields $T_{L,R}^0$ of charge $2/3$ and non-standard isospin assignments are added to the SM. This situation leads to a deviation in their couplings to the $Z$ boson. Atomic parity violation experiments and the measurement of $R_c$ at LEP impose constraints on these deviations for the up and charm quarks, significantly stronger than those for the top quark.
In the Higgs basis, the Yukawa Lagrangian contains the following terms:
\begin{equation}
-\mathcal{L} \,\, \supset  \,\, y^u \bar{Q}^0_L \tilde{H}_2 u^0_R +  y^d \bar{Q}^0_L H_1 d^0_R + M^0_u \bar{u}^0_L u^0_R  + M^0_d \bar{d}^0_L d^0_R + \rm {h.c}.
\end{equation}
Here, $u_R$ actually runs over $(u_R, c_R, t_R, T_R)$ and $d_R$ actually runs over $(d_R, s_R, b_R, B_R)$. 

We now turn to the mixing of the new partners to the third generation, $y_u$ and $y_d$, which are $3\times 4$ Yukawa matrices. In fact, in the light of the above constraints, 
it is very reasonable to assume that only the top quark $t$  ``mixes'' with $T$.
In this case, the $2 \times 2$ unitary matrices $U_{L,R}^u$ define the relation between the charge $2/3$ weak and mass eigenstates:
\begin{eqnarray}
\left(\! \begin{array}{c} t_{L,R} \\ T_{L,R} \end{array} \!\right) &=&
U_{L,R}^u \left(\! \begin{array}{c} t^0_{L,R} \\ T^0_{L,R} \end{array} \!\right)
\nonumber \\
&=& \left(\! \begin{array}{cc} \cos \theta_{L,R}^u & -\sin \theta_{L,R}^u e^{i \phi_u} \\ \sin \theta_{L,R}^u e^{-i \phi_u} & \cos \theta_{L,R}^u \end{array}
\!\right)
\left(\! \begin{array}{c} t^0_{L,R} \\ T^0_{L,R} \end{array} \!\right) \,.
\label{ec:mixu}
\end{eqnarray}
In contrast to the up-type quark sector, the addition of new fields $B_{L,R}^0$ of charge $-1/3$ in the down-type quark sector results in four mass eigenstates $d,s,b,B$.
Measurements of $R_b$ at LEP set constraints on the $b$ mixing with the new fields that are stronger than for mixing with the lighter quarks $d,s$. 
In this case, then,  $2 \times 2$ unitary matrices $U_{L,R}^d$ define the dominant $b-B$ mixing as 
\begin{eqnarray}
\left(\! \begin{array}{c} b_{L,R} \\ B_{L,R} \end{array} \!\right)
&=& U_{L,R}^d \left(\! \begin{array}{c} b^0_{L,R} \\ B^0_{L,R} \end{array} \!\right)
\nonumber \\
&=& \left(\! \begin{array}{cc} \cos \theta_{L,R}^d & -\sin \theta_{L,R}^d e^{i \phi_d} \\ \sin \theta_{L,R}^d e^{-i \phi_d} & \cos \theta_{L,R}^d \end{array}
\!\right)
\left(\! \begin{array}{c} b^0_{L,R} \\ B^0_{L,R} \end{array} \!\right) \,.
\label{ec:mixd}
\end{eqnarray}

(More details on this Lagrangian formalism are shown in the Appendix.) 
To ease the notation, we have dropped the superscripts $u$($d$) whenever the mixing occurs only in the up(down)-type quark sector. 
Additionally, we sometime use the shorthand notations $s_{L,R}^{u,d} \equiv \sin \theta_{L,R}^{u,d}$, $c_{L,R}^{u,d} \equiv \cos \theta_{L,R}^{u,d}$, etc.

This Lagrangian contains all the phenomenological relevant information:
\begin{itemize}
	\item[(i)] the modifications of the SM couplings that might show indirect effects of new quarks can be found in the terms that do not contain heavy quark fields;
	\item[(ii)] the terms relevant for LHC phenomenology (i.e., heavy quark production and decay) are those involving a heavy and a light quark;
	\item[(iii)] terms with two heavy quarks are relevant for their contribution to oblique corrections.
\end{itemize}
In the weak eigenstate basis, the diagonalization of the mass matrices makes the Lagrangian of the third generation and heavy quark mass terms such as

\begin{eqnarray}
\mathcal{L}_\text{mass} & = & - \left(\! \begin{array}{cc} \bar t_L^0 & \bar T_L^0 \end{array} \!\right)
\left(\! \begin{array}{cc} y_{33}^u \frac{v}{\sqrt 2} & y_{34}^u \frac{v}{\sqrt 2} \\ y_{43}^u \frac{v}{\sqrt 2} & M^0 \end{array} \!\right)
\left(\! \begin{array}{c} t^0_R \\ T^0_R \end{array}
\!\right) \notag \\
& & - \left(\! \begin{array}{cc} \bar b_L^0 & \bar B_L^0 \end{array} \!\right)
\left(\! \begin{array}{cc} y_{33}^d \frac{v}{\sqrt 2} & y_{34}^d \frac{v}{\sqrt 2} \\ y_{43}^d \frac{v}{\sqrt 2} & M^0 \end{array} \!\right)
\left(\! \begin{array}{c} b^0_R \\ B^0_R \end{array}
\!\right) +\text{h.c.},
\label{ec:Lmass}
\end{eqnarray}
with $M^0$ a bare mass
term\footnote{As pointed out in the introduction, this bare mass term is not related to the Higgs mechanism. It is gauge-invariant and can appear as such in the Lagrangian, or it can be generated by a Yukawa coupling to a scalar multiplet that acquires a VEV $v' \gg v$.}, $y_{ij}^q$, $q=u,d$, Yukawa couplings and  $v=246$ GeV the Higgs VEV in the SM. Using the standard techniques of diagonalization, the mixing matrices are obtained by
\begin{equation}
U_L^q \, \mathcal{M}^q \, (U_R^q)^\dagger = \mathcal{M}^q_\text{diag} \,,
\label{ec:diag}
\end{equation}
with $\mathcal{M}^q$ the two mass matrices in Eq.~(\ref{ec:Lmass}) and $\mathcal{M}^q_\text{diag}$ the diagonals ones. 
To check the consistency of our calculation, the corresponding $2 \times 2$ mass matrix reduces to the SM quark mass term if either the $T$ or $B$ quarks are absent.

Notice also that, in multiplets with both $T$ and $B$ quarks, the bare mass term is the same for the up-and down-type quark sectors. For singlets and triplets one has, $y_{43}^q = 0$ whereas for doublets $y_{34}^q=0$. Moreover, for the \xtb\ triplet one has $y_{34}^d = \sqrt 2 y_{34}^u$ and for the \tby\ triplet one has $y_{34}^u = \sqrt 2 y_{34}^d$\footnote{We write the triplets in the spherical basis, hence, the $\sqrt 2$ factors stem from the relation between the Cartesian and spherical coordinates of an irreducible tensor operator of rank 1 (vector).}.

The mixing angles in the left- and right-handed sectors are not independent parameters. From the mass matrix bi-unitary diagonalization in Eq.~(\ref{ec:diag}) one finds:
\begin{eqnarray}
\tan 2 \theta_L^q & = & \frac{\sqrt{2} |y_{34}^q| v M^0}{(M^0)^2-|y_{33}^q|^2 v^2/2 - |y_{34}^q|^2 v^2/2} \quad \text{(singlets, triplets)} \,, \notag \\
\tan 2 \theta_R^q & = & \ \frac{\sqrt{2}  |y_{43}^q| v M^0}{(M^0)^2-|y_{33}^q|^2 v^2/2 - |y_{43}^q|^2 v^2/2} \quad \text{(doublets)} \,,
\label{ec:angle1}
\end{eqnarray}

with the relations:
\begin{eqnarray}
\tan \theta_R^q & = & \frac{m_q}{m_Q} \tan \theta_L^q \quad \text{(singlets, triplets)} \,, \notag \\
\tan \theta_L^q & = & \frac{m_q}{m_Q} \tan \theta_R^q \quad \text{(doublets)} \,,
\label{ec:rel-angle1}
\end{eqnarray}
with $(q,m_q,m_Q) = (u,m_t,m_T)$ and $(d,m_b,m_B)$, so one of the mixing angles is always dominant, especially in the down-type quark sector. In addition, for the triplets, the relations between the off-diagonal Yukawa couplings lead to relations between the mixing angles in the up-and down-type quark  sectors,
\begin{eqnarray}
\sin 2\theta_L^d & = & \sqrt{2} \, \frac{m_T^2-m_t^2}{m_B^2-m_b^2} \sin 2 \theta_L^u \quad \quad (X\,T\,B) \,, \notag \\
\sin 2\theta_L^d & = & \frac{1}{\sqrt{2}} \frac{m_T^2-m_t^2}{m_B^2-m_b^2} \sin 2 \theta_L^u \quad \quad (T\,B\,Y) \,.
\label{TBY-mix}
\end{eqnarray}
The masses of the heavy VLQs deviate from $M^0$ due to the non-zero mixing with the SM quarks and for doublets and triplets, the masses of the different components of the multiplet are related. Altogether, these relations show that all multiplets except the \tb\ doublet can be parametrized by a mixing angle, a heavy quark mass and a CP-violating phase that enters some couplings, with the latter being ignored for the observables considered in this paper. In the case of the \tb\ doublet, there are two independent mixing angles and two CP-violating phases for the up-and-down-type quark sectors, with - again - the latter set to zero in our analysis. Hereafter, we refer to such a construct as the 2HDM+VLQ scenario, each distinguishing between the singlet, doublet, and triplet cases. In the present paper, though, given the emphasis on $T$ VLQs, as opposed to $B$ VLQs, we will treat the $(T)$ singlet,   $(XT)$ and $(TB)$ doublets as well as $(XTB)$ and $(TBY)$ triplets whereas we will not deal with the $(B)$ singlet and $(BY)$ doublet representations, as their study is deferred to a future publication. Finally, as discussed in the abstract, we are bound to work in the so-called ``alignment limit'' of the 2HDM \cite{Draper:2020tyq}, wherein we fix $m_h=125$ GeV (so that the lightest neutral Higgs state of the 2HDM is the discovered one) and we have further taken $m^2_{12}=m_A^2\frac{\tan^2\beta}{1+\tan^2\beta}$.

\subsection{Implementation and validation}
In this subsection, we briefly describe our implementation of the aforementioned BSM model. We have used \texttt{2HDMC-1.8.0} \cite{Eriksson:2009ws} as a base platform for our 2HDM+VLQ setup\footnote{A public release of it is in progress:  herein, the analytical expressions for the Feynman rules of the interaction vertices of our 2HDM+VLQ model have been implemented as a new class while several new tree-level VLQ decays have explicitly been coded alongside those of Higgs bosons into VLQs themselves.}.  As a first step, the above Lagrangian components were implemented into \texttt{FeynRules-2.3} \cite{Degrande:2011ua} to generate the proper spectrum of masses and couplings. With the help of this program, we have then generated \texttt{FeynArts-3.11} \cite{Hahn:2000kx,Kublbeck:1990xc} and \texttt{FormCalc-9.10} \cite{Hahn:2001rv,Hahn:1998yk} model files as well as Universal FeynRules Output (UFO) interfaces to be used in\\ \texttt{MadGraph-3.4.2} \cite{Alwall:2014hca}. As consistency checks, we have verified the cancellation of Ultra-Violet (UV)  divergences as well as the renormalization scale independence for loop-level processes ${h,H,A} \to \gamma\gamma, \gamma Z$, and $gg$.

\subsection{EWPOs in VLQ}
EWPOs can restrict severely the parameter space 
of BSM physics scenarios. In our 2HDM+VLQ setup,  for both $S$ and $T$, we have two additive contributions: one from VLQs and the other from 2HDM Higgses. The latter is very well known, and we have taken it from \cite{Kanemura:2015mxa} while the former can be calculated using the approach of  Ref.~\cite{Lavoura:1992np}, as it has been done in  Ref.~\cite{Chen:2017hak}. However, the approach of  Ref.~\cite{Lavoura:1992np} works only for singlet and doublet representation of the new fermions, and cannot be applied to the triplet cases.
For all various VLQs models considered here, we have computed the $S$ and $T$ parameters with the use of the aforementioned \texttt{FeynArts} and \texttt{FormCalc} packages  using dimensional regularization. The analytic results were obtained as a combination of standard Passarino-Veltman functions. We checked both analytically and numerically that our results for $S$ and $T$ are UV finite and renormalization scale independent. 
(Such analytic results for the $S$ and $T$ parameters will be presented  in \cite{Abouabid:2023mbu}.) For the $T$  parameter, we have compared our results to  Refs.~\cite{Lavoura:1992np,Chen:2017hak} and found good agreement in the case of singlet and doublet. In the case of the triplet representations $(XTB)$ and $(TBY)$, our result disagrees with Ref.~\cite{Chen:2017hak}. Furthermore, for the $S$ parameter,  our results disagree with those of Ref. \cite{Chen:2017hak}. In fact, Ref.~\cite{Chen:2017hak}  simply extends the result of the singlet and doublet cases from Ref.~\cite{Lavoura:1992np}  to the triplet one. The author of  Ref.~\cite{He:2022zjz}  derives the correct expression for $S$ and $T$ in the case of the  $(XTB)$ triplet using the approach of  Ref.~\cite{Lavoura:1992np}.  We re-computed from scratch the results for the $(XTB)$ and $(TBY)$ triplet cases, then re-derived $S$ and $T$ in terms of Passarino-Veltman functions and, crucially, did not neglect the external momentum of the gauge bosons.
We are confident of the correctness of our findings, as they were derived in two different ways. A major source of disagreement would be the fact that Ref.~\cite{Chen:2017hak} uses a simple approximation for the $S$ parameter, which consists in neglecting the external momentum of the gauge bosons. We also cross-checked with a new calculation \cite{Cao:2022mif,He:2022zjz} and found good agreement for the $T$ parameter. However, we should mention that, for the $S$ parameter, we use the complete analytical result and, again, did not neglect any external momentum as it is usually the case \cite{Cao:2022mif}. 
\section{Constraints}
\label{sec-A}
In this section, we list the constraints that we have used to check the validity of our results.
From the theoretical side, we have the following requirements:
\begin{itemize}
	\item \textbf{Unitarity} constraints require the $S$-wave component of the various
	(pseudo)scalar-(pseudo)scalar, (pseudo)scalar-gauge boson, and gauge-gauge bosons scatterings to be unitary
	at high energy ~\cite{Kanemura:1993hm}.
	\item \textbf{Perturbativity} constraints impose the following condition on the quartic couplings of the scalar potential: $|\lambda_i|<8\pi$ ($i=1, ...5$)~\cite{Branco:2011iw}.    
	\item \textbf{Vacuum stability} constraints require the potential to be bounded from below and positive in any arbitrary
	direction in the field space, as a consequence, the $\lambda_i$ parameters should satisfy the conditions as~\cite{Barroso:2013awa,Deshpande:1977rw}:
	\begin{align}
	\lambda_1 > 0,\quad\lambda_2>0, \quad\lambda_3>-\sqrt{\lambda_1\lambda_2} ,\nonumber\\ \lambda_3+\lambda_4-|\lambda_5|>-\sqrt{\lambda_1\lambda_2}.\hspace{0.5cm}
	\end{align} 
	\item \textbf{Constraints from EWPOs}, implemented through the oblique parameters, $S$ and $T$ ~\cite{Grimus:2007if},  require that, for a parameter point of our
	model to be allowed, the corresponding $\chi^2(S^{\mathrm{2HDM\text{-}II}}+S^{\mathrm{VLQ}},~T^{\mathrm{2HDM\text{-}II}}+T^{\mathrm{VLQ}})$ is within 95.45\% Confidence Level (CL) in matching the global fit results \cite{ParticleDataGroup:2020ssz}:
	\begin{align}
	&S= 0.05 \pm 0.08,\quad T = 0.09 \pm 0.07,\nonumber\\& \rho_{S,T} = 0.92 \pm 0.11.\hspace{0.5cm}(\text{For~~}U=0) 
	\end{align}
	Note that unitarity, perturbativity, vacuum stability, as well as $S$ and $T$ constraints, are enforced through   \\\texttt{2HDMC-1.8.0}.
\end{itemize}
From the experimental side, we evaluated the following:
\begin{itemize}
	\item \textbf{Constraints from the SM-like Higgs-boson properties}  are taken into account by using \texttt{HiggsSignal-3} \cite{Bechtle:2020pkv,Bechtle:2020uwn} via \texttt{HiggsTools} \cite{Bahl:2022igd}. We require that the relevant quantities (signal strengths, etc.) satisfy $\Delta\chi^2=\chi^2-\chi^2_{\mathrm{min}}$ for these measurements at 95.45\% CL ($\Delta\chi^2\le6.18$), involving 159 observables.
	\item\textbf{Constraints from direct searches at colliders}, i.e., LEP, Tevatron, and LHC, are taken at the 95\% CL and are tested using \texttt{HiggsBouns-6}\cite{Bechtle:2008jh,Bechtle:2011sb,Bechtle:2013wla,Bechtle:2015pma} via \texttt{HiggsTools}. Including the most recent searches for neutral and charged scalars. The principal search channels relevant to the Type-II 2HDM are:

	\begin{itemize}
		\item $p p \rightarrow A \rightarrow Z Z \rightarrow l^+ l^+ l^- l^-,\;l^+l^-qq,\;l^+l^- \nu \bar{\nu} $ \cite{CMS:2018amk}
		\item $b b \rightarrow A \rightarrow Z h \rightarrow l^+ l^- b b $ \cite{CMS:2019qcx}
		\item $p p \rightarrow H/A \rightarrow \tau^+ \tau^- $ \cite{ATLAS:2020zms}
		\item $p p \rightarrow H \rightarrow Z Z \rightarrow l^+l^-l^+l^-, l^+l^-qq,l^+l^-\nu\bar{\nu} $\cite{CMS:2018amk}
		\item $p p \rightarrow H \rightarrow V V $\cite{ATLAS:2018sbw}
		\item $p p \rightarrow A \rightarrow H Z \rightarrow b\bar{b}l^+l^- $\cite{ATLAS:2018oht}
		\item $p p \rightarrow H \rightarrow A Z \rightarrow b\bar{b}l^+l^- $\cite{ATLAS:2018oht}
		\item $ g g \rightarrow A \rightarrow h Z \rightarrow b\bar{b}l^+l^- $\cite{ATLAS:2018oht}
		\item $p p \rightarrow tbH^+ \rightarrow tbtb$ \cite{ATLAS:2021upq}
	\end{itemize}
	
	\item {\bf Constraints from $\boldsymbol{b\to s\gamma}$}: To align with the $b\to s\gamma$ limit, the mass of the charged Higgs boson is set to exceed 600 GeV. This is based on the analysis in Ref \cite{Benbrik:2022kpo}, which indicates that incorporating VLQs into the Type-II 2HDM could relax this limit through larger mixing angles. However, EWPOs constrain these angles to smaller values, consequently maintaining the charged Higgs mass close to the typical 580 GeV in the 2HDM Type-II.		
\end{itemize}

\subsection{Direct search constraints}

\begin{figure}[H]
	\centering
	\includegraphics[width=0.45\textwidth,height=0.42\textwidth]{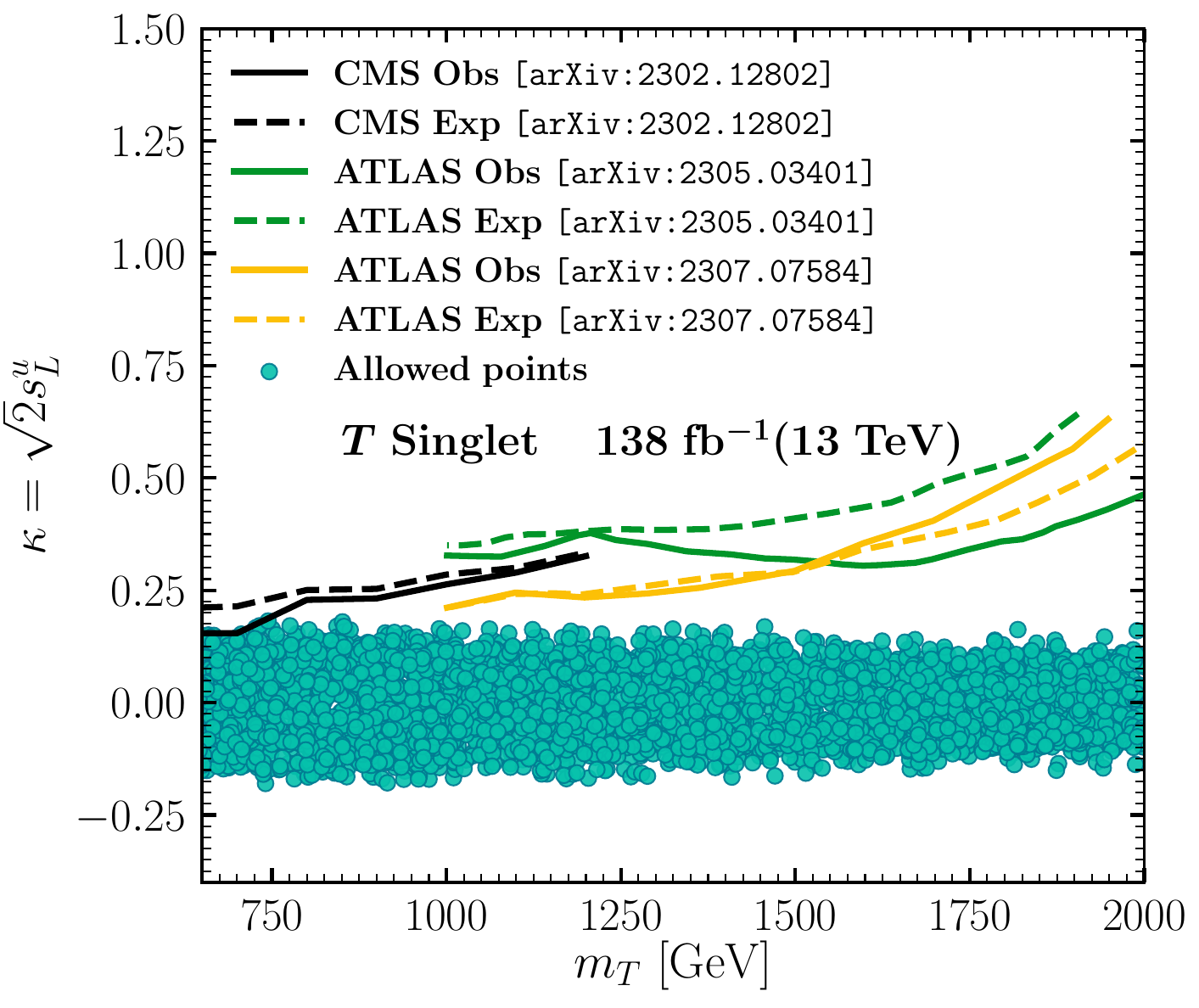}
	\caption{Allowed points following the discussed theoretical and experimental constraints,in the ($m_T, \kappa$) plane for the 2HDM+$T$ singlet scenario, superimposed onto the CMS \cite{CMS:2023agg} (black) and ATLAS \cite{ATLAS:2023pja,ATLAS:2023bfh} (green/yellow) 95\% C.L observed (solid line) and expected (dashed) upper limits on the coupling $\kappa$.}
	\label{fig_1}
\end{figure}

While the experimental constraints on the properties (masses and couplings) of additional Higgs states in the 2HDM+VLQ framework can be efficiently managed with the described toolbox, the limits on VLQs require a dedicated analysis. Our study demonstrates that the oblique parameters $S$ and $T$ impose stringent restrictions on VLQ mixing angles, resulting in small values that maintain compatibility with current LHC bounds on VLQ masses. This allows us to explore a broad mass range beginning at 600 GeV, as shown in Fig.~\ref{fig_1}. Although the presented example focuses on the singlet scenario, similar behaviour is observed in the doublet and triplet cases.
Fig.~\ref{fig_1} presents the allowed points for the 2HDM+$T$ singlet scenario in the $(m_T, \kappa)$ plane, overlaid with the 95\% Confidence Level (CL) observed and expected limits from CMS \cite{CMS:2023agg} and ATLAS \cite{ATLAS:2023pja,ATLAS:2023bfh}. The plot shows that, under the constraints from the $S$ and $T$ parameters, almost all\footnote{Our analysis distinguishes data points based on their mass, particularly those above and below $m_T = 1000$ GeV. Data points exceeding 1000 GeV lie well within the 95\% CL limit, while points in the 600--800 GeV range approach the exclusion line. To ensure alignment with these limits, we incorporated these criteria directly into our code, excluding points that exceed the 95\% CL.} points remain within the observed 95\% CL exclusion limits, supporting VLQ masses as low as 600 GeV within this framework.

For VLQ masses in the range of 600--1000 GeV, small mixing angles (e.g., $\sin\theta \lesssim 0.12$) ensure consistency with current experimental limits. The constraints shown in Fig.~\ref{fig_1} primarily arise from single production searches, which depend on electroweak interactions and are sensitive to coupling strength. By contrast, constraints derived from the pair production mode of VLQs are generally more restrictive, excluding singlet $T$ masses below roughly 1480 GeV \cite{CMS:2022fck} and up to 1700 GeV when ${\cal BR}(T \to Wb) = 100\%$ \cite{ATLAS:2024gyc}. In the 2HDM+VLQ framework, however, the inclusion of additional non-SM decay channels (e.g., $T \to Ht$, $T \to At$) reduces the branching ratios into SM states, thereby relaxing these exclusion limits. Notably, in our analysis, these constraints are applied strictly only when the branching ratio into non-SM channels, ${\cal BR}(T \to \text{non-SM})$, is zero.

\section{VLQs contribution to Higgs production and decay}

VLQs contribute to the loop diagrams for Higgs production via gluon-gluon fusion and Higgs decay into two photons. However, in our study, their contributions are relatively small. This is partly because VLQs tend to decouple from the process as their masses increase and also due to the mixing angles terms $({s^{u}_{L,R}})^2$ appear in the coupling $hT\overline T$  (see Table XIV of the appendix) are restricted by the EWPOs resulting in small values for these terms (The largest allowed mixing angle is approximately $s^u_{L,R}\sim 0.2$), further suppressed by their squares. However, it's noteworthy that the primary, albeit small, contribution arises from modifications in the $ht\bar t$ coupling.

For the processes $gg\to h$ (or $h \to gg$) and $h \to \gamma \gamma$, the contribution of all quarks of the same charge to the corresponding amplitudes is determined by the expression\footnote{The LHC searches relevant to the $h\to \gamma \gamma$ process, as incorporated in \texttt{HiggsSignals-3}, are  \cite{ATLAS:2022tnm,CMS:2021kom}.}:
\begin{equation}
F_q=\sum_i Y_{ii} A_{1/2}\left(\frac{M_H^2}{4 m_i^2} \right),
\end{equation}
Here, the sum runs over $t,T$ for $q=u$ and over $b,B$ for $q=d$. The couplings to the Higgs $Y_{ii}$ are defined in Table XIV of the appendix, and the function $A_{1/2}$ is defined in~\cite{Djouadi:2005gi}.

Our analysis reveals that, across all considered scenarios, the branching ratio $\mathcal{BR}(h\to gg)$ experiences a decrease of up to 10\%, primarily due to modifications in the $ht \bar t$ coupling. Similarly, the branching ratio $\mathcal{BR}(h\to \gamma\gamma)$ decreases by up to 3\%, also attributed to modifications in the $ht\bar t$ coupling. These reductions are well within the current experimental precision limits \cite{ATLAS:2021vrm}.  Contributions from new terms such as $h T \overline{T}$ are considered negligible, as discussed previously.

\section{Results and discussions}
\subsection{2HDM with $(T)$ singlet}
\begin{figure*}[h!]
	\centering
	\includegraphics[width=0.85\textwidth,height=0.425\textwidth]{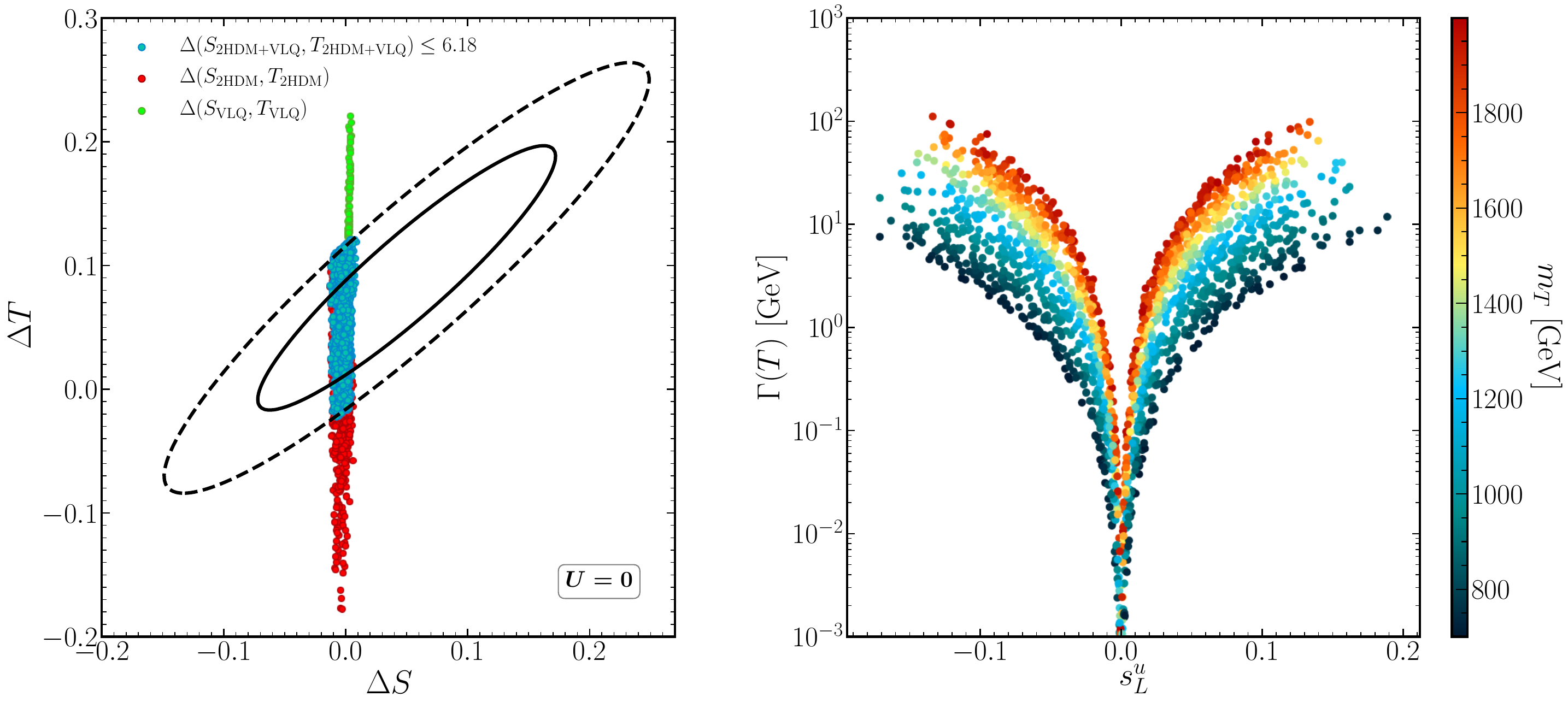}
	\caption{(Left) Scatter plot of randomly generated points superimposed on the fit limits in the $(\Delta S, \Delta T)$ plane from EWPO data at the 95.45\% confidence level, with a correlation of 92\% \cite{ParticleDataGroup:2020ssz}. The black solid line marks the 1$\sigma$ CL region for $\chi_{ST}^2$, while the dashed line denotes the 2$\sigma$ CL region under the assumption $U=0$. Separate and combined contributions from the 2HDM and VLQ are displayed. (Right) The $T$ width $\Gamma(T)(\equiv \Gamma_T)$ as a function of $s_L^u\equiv\sin\theta_L$ with $m_T$ indicated as a colour gauge.}

	\label{fig1}
\end{figure*}

\begin{table*}[t!]
	\centering
	{\renewcommand{\arraystretch}{1} 
		{\setlength{\tabcolsep}{2.5cm}
			\begin{tabular}{c  c}
				\toprule\toprule
				Parameters  & Scanned ranges \\
				\toprule			
				$m_h$   & $125$ \\
				$m_A$  & [$200$, $1000$] \\
				$m_H$  & [$200$, $1000$] \\
				$m_{H^\pm}$  & [$600$, $1000$] \\
				$\tan\beta$ & [$1$, $20$] \\
				$m_{T}$   & [$650$, $2000$] \\	
				$\sin\theta_L^{u,d}$  & [$-0.8$, $0.8$] \\
				$\sin\theta_R^{u,d}$  & [$-0.8$, $0.8$] \\
				\toprule\toprule
	\end{tabular}}}
	\caption{2HDM and VLQs parameters for all scenarios with their scanned ranges. Masses are in GeV. Phases $\phi_u$ and $\phi_d$ are set to zero.}
	\label{table1}
\end{table*}
In Fig.~\ref{fig1} we perform a scan over the 2HDM  parameters (the Higgs masses, $\tan\beta$, $\sin(\beta-\alpha)$, $m_{12}$) plus the singlet top mass $m_T$ and the (fermionic) mixing angle $\sin\theta_L$,  as indicated in Tab.~\ref{table1}. In  Fig.~\ref{fig1} (left), we illustrate the results in terms of the contribution of the 2HDM scalars as well as of the only VLQ of this 2HDM+VLQ scenario to the $S$ and $T$ parameters. We do so by showing the two contributions separately as well as summed together. 
The individual $S_{\rm VLQ,\; 2HDM}$ terms are rather small while the $T_{\rm VLQ,\; 2HDM}$ ones  can be large and with opposite
sign, thus allowing for strong cancellations, in particular, $T_{\rm 2HDM}$ can have both signs while $T_{\rm VLQ}$ is (nearly) always positive. A large and positive $T_{\rm VLQ}$ is possible for large, $\sin\theta_L$ while a large and negative $T_{\rm 2HDM}$ is possible in the 2HDM with large splitting among heavy Higgs masses. Note that, in this multiplet case, the EWPO constraints from $S$ and $T$ are stronger than $R_b$ limits \cite{Aguilar-Saavedra:2013qpa}. 

In the right panel of Fig.~\ref{fig1}, {{we project our scan onto the plane $(\sin\theta_L,\Gamma_T)$, where one can see that the mixing angle $\mid\sin\theta_L\mid$ is typically constrained to be less than 0.20 for $m_T$ around 700--800 GeV. This value is reduced to approximately 0.15 for $m_T=2$ TeV.  For small mixing, $\mid\sin\theta_L\mid<0.05$, the total width is rather small irrespectively of $m_T$, of the order few GeV or (much) less, while for large mixing,  $\mid\sin\theta_L\mid \approx 0.05-0.15$,  and large  $m_T$, the total width can be in the range $10-100$ GeV.  Hence, it is to be noted that such $\Gamma_T$ values will probably be comparable with the experimental resolution of reconstructed $T$ states, a regime in which phenomenological implications have been studied in \cite{Moretti:2016gkr,Carvalho:2018jkq} (see also \cite{Prager:2017hnt,Prager:2017owg,Moretti:2017qby}). In the end, the $T$ state width can reach at most the 10\% level of its mass. 
		
		Altogether, then,  the presence of the additional degrees of freedom of the VLQ sector, combined with those from the 2HDM one, allows for solutions covering rather light $m_T$ values and rather large mixing $|\sin\theta_L|$, to potentially enable significant $pp\to T\bar T$ production rates at the LHC as well as $T$ decays into heavy (pseudo)scalars Higgs states of 2HDM origin with large probabilities.}} 

We now move on to discuss each Branching Ratio (${\cal BR}$) of the  $T$ state. In the SM with an additional singlet top, the picture is rather simple. The ${\cal BR}$'s  of $T\to W^+b$, $T\to Zt$ and $T\to ht$ are approximately around  50\%, 25\% and   25\%, respectively. These three ${\cal BR}$'s are not very sensitive to the mixing angle $\sin\theta_L$ and, further,  are not independent of one another, as they satisfy the following sum rule:
\begin{eqnarray}
{\cal BR}(T\to {\rm SM}) &=& {\cal BR}(T\to W^+b)\nonumber,\\
&+& {\cal BR}(T\to Zt)+ {\cal BR}(T\to ht)=1.
\label{eq:sumrule1} 
\end{eqnarray}
(Here, we neglect the decays that are either Cabibbo-Kobayashi-Maskawa (CKM) suppressed or that are loop-mediated, such as $T\to c g$, $T\to c\gamma$, etc.) When we add extra (pseudo)scalars, the picture changes drastically because of the presence of new decay channels, such as $T\to H^+b$, $T\to At$ and $T\to Ht$,  that could significantly modify the limits coming from $T$  direct searches at the LHC (as explained).  The above sum rule will be modified as follows:
\begin{eqnarray}
&&{\cal BR}(T\to {\rm SM}) + {\cal BR}(T\to {\rm non~SM}) =1,\nonumber \\
 &&{\cal BR}(T\to {\rm non~SM}) = {\cal BR}(T\to H^+b)+ {\cal BR}(T\to At)\nonumber \\
&&\qquad\qquad\qquad\qquad+ {\cal BR}(T\to Ht).
\label{eq:sumrule2} 
\end{eqnarray}
\begin{figure*}[t!]
	\centering
	\includegraphics[width=1\textwidth,height=0.40\textwidth]{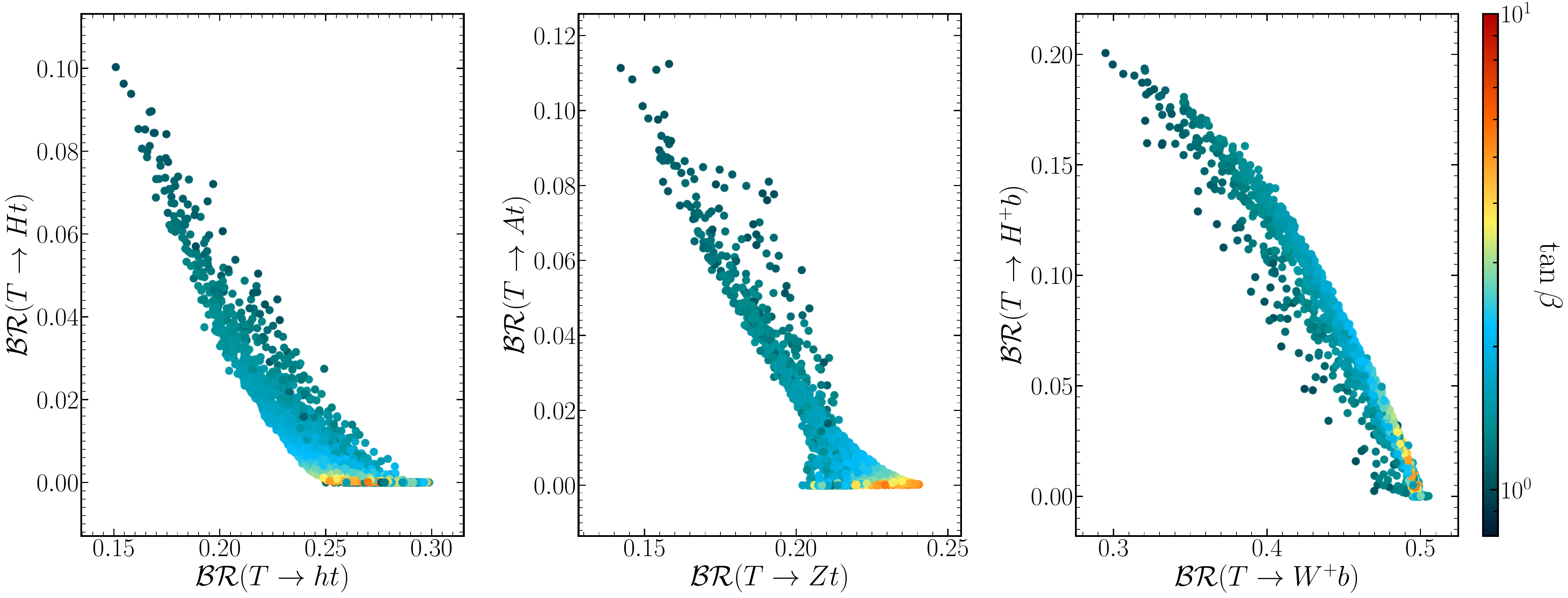}
	\caption{The correlation between ${\cal BR}(T\to ht)$ and ${\cal BR}(T\to Ht)$ (left), ${\cal BR}(T\to Zt)$ and ${\cal BR}(T\to At)$ (middle) as well as  ${\cal BR}(T\to W^+b)$ and ${\cal BR}(T\to H^+b)$ (right) with $\tan\beta$ indicated in the colour gauge.}	
	\label{fig2}
\end{figure*}

\begin{figure*}[t!]
	\centering
	\includegraphics[width=0.85\textwidth,height=0.425\textwidth]{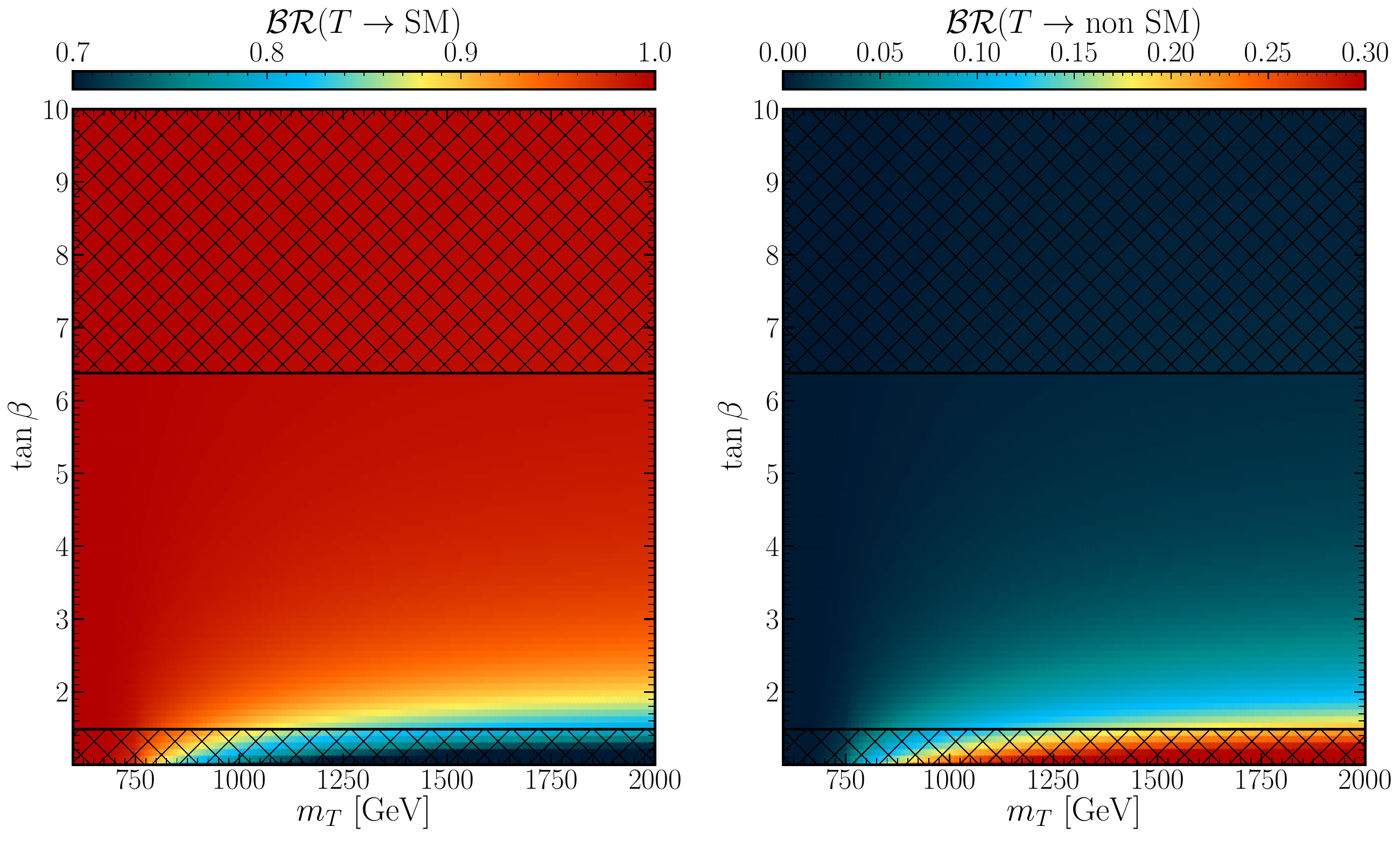}
	\caption{The ${\cal BR}(T\to$ SM) (left) and ${\cal BR}(T\to$ non SM)  (right) mapped onto the $(m_T, \tan\beta)$ plane, with $\sin\theta_L=0.045$, $\sin(\beta-\alpha)=1$, $m_h=125$ GeV, $m_H=585$, $m_A=582$ GeV, and $m_{H^\pm}=650$ GeV (recall that $m_{12}^2=m_A^2\tan\beta/(1+\tan^2\beta)$).  Here, the shaded areas are excluded by \texttt{HiggsBounds} ($H^+\to t\bar{b}$ \cite{ATLAS:2021upq} for $\tan\beta<2$ and $H\to\tau\tau$ \cite{ATLAS:2020zms} for $\tan\beta>6$). All other constraints ($S$, $T$, \texttt{HiggsSignals} and theoretical ones) are also checked.}	
	\label{fig3}
\end{figure*}
\begin{table*}[h!]
	\begin{center}
		\setlength{\tabcolsep}{45pt}
		\renewcommand{\arraystretch}{1}
		\begin{adjustbox}{max width=\textwidth}		
			\begin{tabular}{lcc}
				\toprule\toprule
				Parameters &       BP$_1$ &       BP$_2$ \\
				\toprule

				\multicolumn{3}{c}{2HDM+VLQ inputs. The masses are in GeV.} \\\toprule
				$m_h$   &   125&   125  \\
				$m_H$  &    768.68 &  958.33\\
				$m_A$   & 671.47 &  759.01 \\
				$m_{H\pm}$   &739.59 &  788.66  \\
				$\tan\beta$ &    1.17 &    1.04 \\
				
				$m_T$      &1092.37 & 1931.30 \\
				$\sin(\theta^u)_L$    &   -0.14 &    0.10 \\
				
				\toprule
				\multicolumn{3}{c}{$\mathcal{BR}(H^\pm\to {XY})$ in \%} \\\toprule
				${\cal BR}(H^+\to t\bar{b})$     & 99.81 & 99.82 \\
				${\cal BR}(H^+\to\tau\nu)$ & 0.01 &  0.01 \\
				${\cal BR}(H^+\to W^+ A)$ &  0.01 &  0.00\\
				\toprule
				\multicolumn{3}{c}{$\mathcal{BR}(T\to {XY})$ in \%} \\\toprule
				${\cal BR}(T\to W^+b)$  &   42.38 & 31.31 \\
				${\cal BR}(T\to W^+B)$  &   - &   - \\
				${\cal BR}(T\to Zt)$  &   19.27 & 15.12 \\
				${\cal BR}(T\to ht)$  &   22.73 & 15.97   \\
				${\cal BR}(T\to Ht)$  &    3.03 &  7.99   \\
				${\cal BR}(T\to At)$  &   3.52 &  9.32 \\
				${\cal BR}(T\to H^+b)$ &  9.07 & 20.29 \\
				${\cal BR}(T\to H^+B)$ &   - &  -   \\				
				\toprule
				\multicolumn{3}{c}{Total decay width in GeV.} \\\toprule
				$\Gamma(T)$ & 19.00 &   82.65 \\
				\toprule
				\multicolumn{3}{c}{Observables} \\\toprule
				$T_{\mathrm{2HDM}}$  &  -0.0353 &   -0.0892 \\
				$T_{\mathrm{VLQ}}$  &  0.1442 &    0.1219 \\
				$S_{\mathrm{2HDM}}$ &  -0.0013 &    0.0046\\
				$S_{\mathrm{VLQ}}$ &    0.0025 &    0.0027   \\	
				$\Delta\chi^2(S_{\mathrm{2HDM+VLQ}},T_{\mathrm{2HDM+VLQ}})$ &   4.87 &    0.98\\\toprule
				$\chi^2{(h_{125})}\equiv \chi^2_{\texttt{HiggsSignals}}$ &  158.49 &  158.58  \\	
				
				\bottomrule\bottomrule
				
			\end{tabular}
		\end{adjustbox}
	\end{center}
	\caption{The full description of our BPs for the $(T)$ singlet case.}\label{Bp1}
\end{table*}
In  Fig.~\ref{fig2},  we illustrate the correlation between the decays of  the $T$ state into SM particles and  those via  the new channels, paired as follows:   
$T\to \{ht, Ht\}$ (left),  $T\to \{Zt,At\}$ (middle) and $T\to \{W^+b,H^+b\}$ (right). 
We note that, once the non-SM decays are kinematically open,  ${\cal BR}(T\to H^+b$) would compete with ${\cal BR}(T \to W^+b)$.  This is also true for ${\cal BR}(T\to Ht)$ and ${\cal BR}(T\to At)$.  

In understanding the numerical results, it is useful to look at the analytical formulae of the Higgs couplings to SM quarks and VLQs\footnote{It is important to note that in this study, we focus solely on the on-shell decays of  $T$ i.e $m_T> m_{\mathrm{Higgs}}$.} (specific to the singlet case), which, in the exact alignment limit, take the following form (in Type-II):
\begin{eqnarray}
&&hTt= \frac{g}{2M_W} h \bar{t}( \cot\beta m_t c_L s_L P_L + m_T c_Ls_L P_R)T\nonumber,\\
&&HTt= \frac{g}{2M_W} H \bar{t}( \cot\beta m_t c_L s_L P_L + \cot\beta m_T c_Ls_L P_R)T\nonumber,\\
&&ATt= \frac{g}{2M_W} A \bar{t}( \cot\beta m_t c_L s_L P_L + \cot\beta m_T c_Ls_L P_R)T\nonumber,\\
&&H^+\overline{T}b= -\frac{gm_T}{\sqrt{2}M_W} H^+ \overline{T}( \cot\beta s_L P_L +\frac{m_b}{m_T}\tan\beta P_R s_L)b\nonumber.
\label{hTt:coup}
\end{eqnarray}
By comparing these formulae to the corresponding ones in the Appendix (Tabs. XVIII--XX, where the full dependence on  $\alpha$ and $\beta$ is retained), we start by noting that, in the alignment limit $\cos(\beta-\alpha)\to 0$, the $\cot\beta m_t$ term in the $hTt$ coupling comes with a suppression factor  $\cos(\beta-\alpha)\approx 0$ while the $HTt$ coupling would come with $\sin(\beta-\alpha)\approx 1$. In contrast, in the case of $ATt$, there is no $\cos(\beta-\alpha)$ term involved. Further, as one can see from the above couplings, in the case of small $\tan\beta\approx 1$, the right-handed couplings of $HTt$ and $ATt$ would get more amplification compared to the left-handed ones because of the term $\cot\beta m_T$  and the fact that $m_T\gg m_t$.  As for the $H^+Tb$ coupling, which has only a left-handed component, this will also enjoy an enhancement for small $\tan\beta\approx 1$. 

In the light of this, examining Fig.~\ref{fig2} (right panel) reveals that at low $\tan\beta$, the branching ratio ${\cal BR}(T\to H^+b)$ reaches up to 20\%, indicating its notable but not dominant size compared to ${\cal BR}(T\to W^+b)$. In contrast,  at large $\tan\beta$, we have the opposite situation, ${\cal BR}(T\to H^+b)$ is quite small while ${\cal BR}(T\to W^+b)$ gets close to its SM value.
Both effects are due to the singlet heavy top coupling to charged Higgs and bottom quark states, which contains a term $\propto m_T/\tan\beta$. We further stress that,  for $\tan\beta \approx 1$, the branching ratios ${\cal BR}$ for the exotic decays $T\to At$, and $T\to Ht$  could reach their maximum values of approximately  11.5\%, and 10\% respectively.

In Fig.~\ref{fig3},  we illustrate the $T$ state cumulative ${\cal BR}$'s into SM particles (left panel)
and non-SM particles (right panel) as a function of $m_T$ and $\tan\beta$. In both cases, the decay of the heavy singlet top partner is not very sensitive to the first parameter. Further, for small $\tan\beta$, the decays into non-SM particles get amplified as $m_T$ grows and reaches a maximum value of 20\% within the permissible region. Conversely, for large $\tan\beta$, irrespectively of $m_T$,  it is the opposite, i.e., the decay into SM particles can become fully dominant.

Thus, having this picture in mind, we initially propose three\footnote{It should be noted that the presentation of all three BPs depends on the fulfilment of these conditions. In scenarios where not all conditions are met, two BPs may be presented.} Benchmark Points (BPs) which are suitable to explore the Type-II scenario of the 2HDM in the presence of a singlet VLQ with top Electromagnetic (EM) charge, as follows.
\begin{itemize} 
	\item[ i)] BP$_1$: where ${\cal BR}(T\to$ SM) is similar to
	${\cal BR}(T\to$ non SM).
	\item[ii)] BP$_2$: where ${\cal BR}(T\to$ SM) is rather small, which makes 
	${\cal BR}(T\to$ non SM) substantial.
	\item[iii)] BP$_3$: where ${\cal BR}(T\to$ SM) is rather large, which makes 
	${\cal BR}(T\to$ non SM) marginal.  
\end{itemize}    
Input parameters for these BPs are given in Tab.~\ref{Bp1}. 


\subsection{2HDM with $(TB)$ doublet}
\begin{figure*}[htpb!]
	\centering
	\includegraphics[width=0.85\textwidth,height=0.425\textwidth]{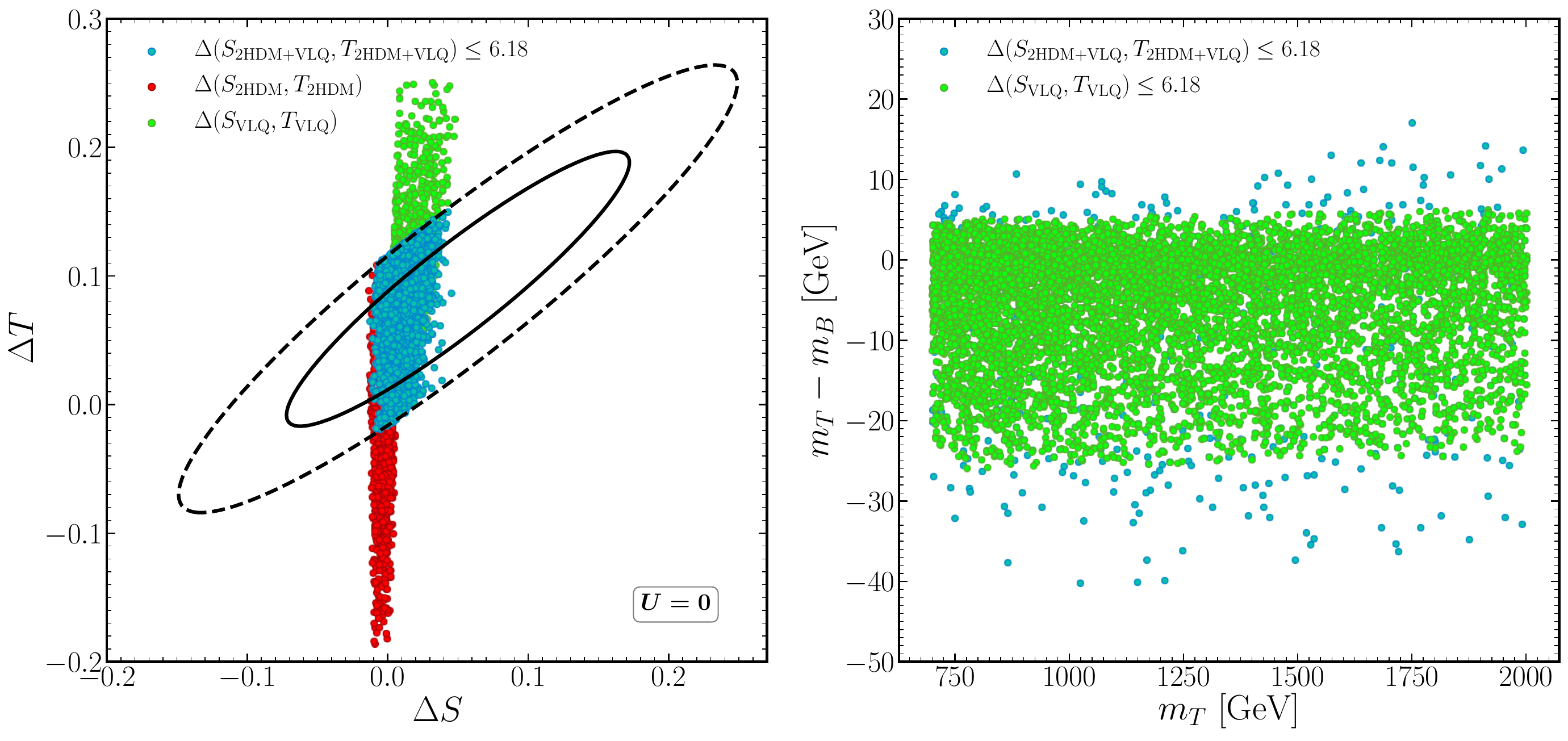} 	
	\caption{(Left) Scatter plot of randomly generated points superimposed on the fit limits in the $(\Delta S, \Delta T)$ plane from EWPO data at the 95.45\% confidence level, with a correlation of 92\% \cite{ParticleDataGroup:2020ssz}. The black solid line marks the 1$\sigma$ CL region for $\chi_{ST}^2$, while the dashed line denotes the 2$\sigma$ CL region under the assumption $U=0$. Separate and combined contributions from the 2HDM and VLQ are displayed. 
		(Right) The same points are mapped onto the $(m_T,\delta)$ plane, where $\delta$ is the mass difference between $T$ and $B$. Here, we only present the VLQ contribution and the total one.  Further, all constraints have been taken into account.} 	
	\label{fig4}
\end{figure*}

\begin{figure*}[htpb!]
	\centering
	\includegraphics[width=0.85\textwidth,height=0.425\textwidth]{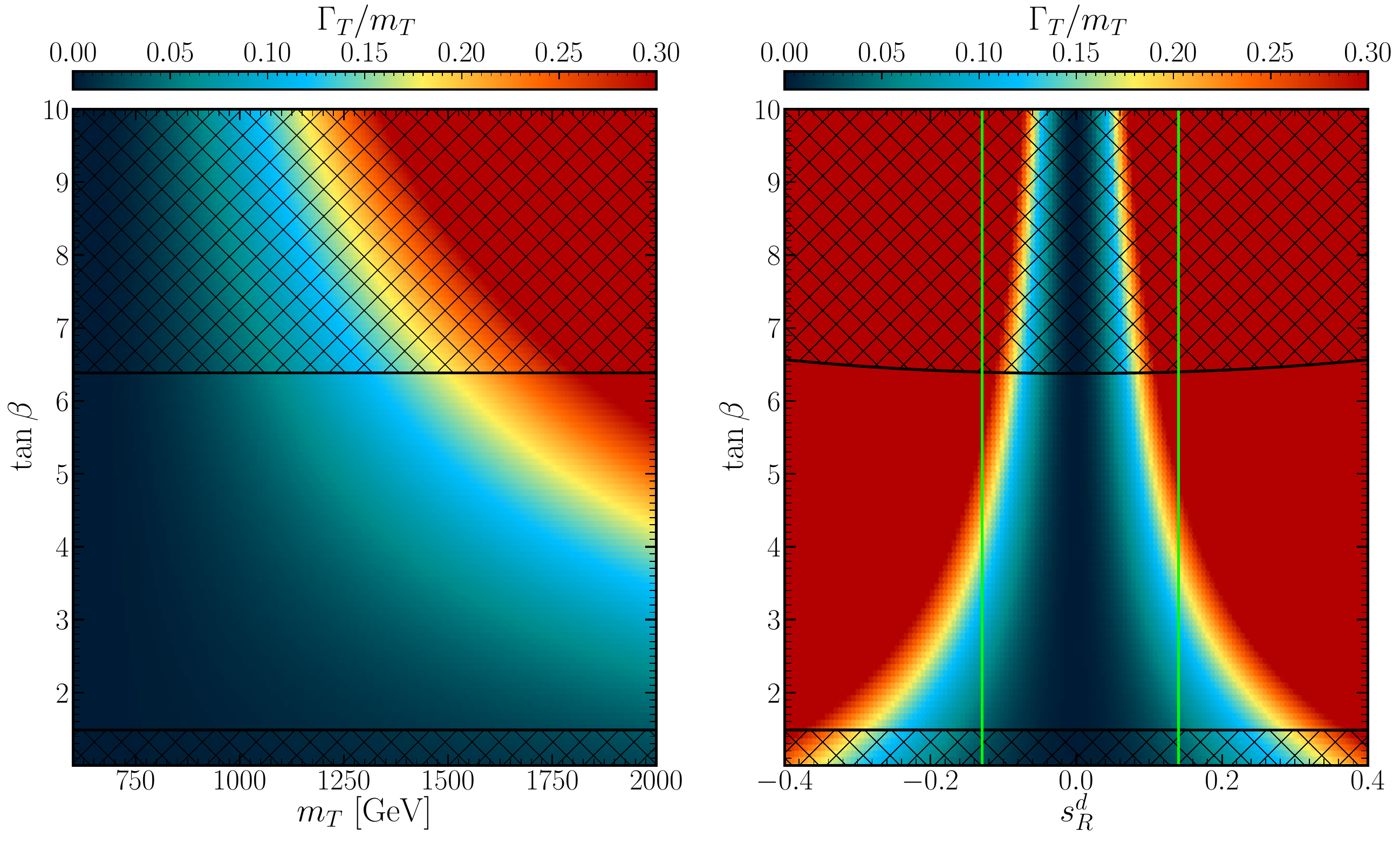}
	\caption{The $\Gamma(T)/m_T$ ratio $(\Gamma(T)\equiv\Gamma_T)$ mapped over the $(m_T,\tan\beta)$ plane (left) and $(s^d_R\equiv\sin\theta^d_R,\tan\beta)$ plane (right), with $\sin\theta^u_R=0.042$ in the left panel and $m_T=1600$ GeV in the right panel (the 2HDM parameters are the same as in Fig. \ref{fig3}). Here, the shaded areas are excluded by \texttt{HiggsBounds}. The regions between the vertical lime green lines are allowed by the $S$, $T$ parameter constraints, all other constraints  (\texttt{HiggsSignals} and theoretical ones) are also checked.}	
	\label{fig5}
\end{figure*}
We now discuss the case of the $(TB)$ doublet.  In the SM extended with such a VLQ multiplet,  both mixing angles in the up-and down-type quark sectors enter the phenomenology of the model\footnote{Hereafter, we replace the $d$ and $u$ superscripts in the mixing angles with $b$ and $t$, respectively.}. 
For  given $\theta_R^b$, $\theta_R^t$ and $m_T$ mass, the relationship between the mass eigenstates  and the mixing angles reads as \cite{Aguilar-Saavedra:2013qpa}:
\begin{eqnarray}
&&m_B^2= (m_T^2 \cos^2\theta_R^{t}+m_t^2 \sin^2\theta_R^{t}-
m_b^2 \sin^2\theta_R^{b})/ \cos^2\theta_R^{b},
\label{tb-mass}
\end{eqnarray}
from where one can compute  the
$m_B$ value. Furthermore,  using Eqs.~(\ref{ec:angle1})--(\ref{ec:rel-angle1}), one can also compute the left mixing $\theta_L^b$ and $\theta_L^t$.

\begin{figure*}[htpb!]
	\centering
	\includegraphics[width=0.85\textwidth,height=0.425\textwidth]{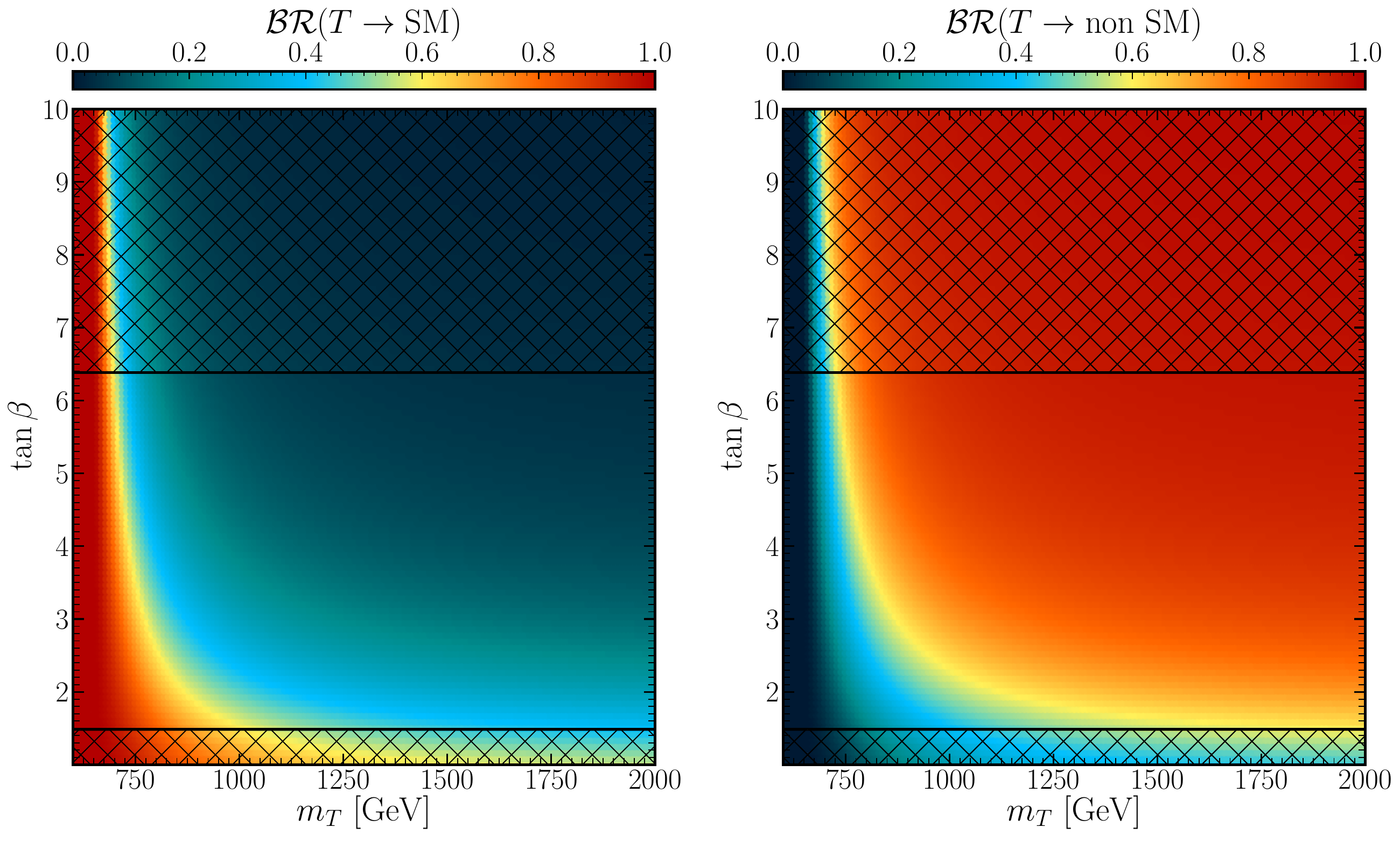}
	\caption{The ${\cal BR}(T\to$ SM) (left) and ${\cal BR}(T\to$ non SM)  (right) mapped onto the $(m_T, \tan\beta)$ plane, with the same description as in Fig.~\ref{fig5} (left).}	
	\label{fig6}
\end{figure*}

\begin{figure*}[htpb!]
	\centering
	\includegraphics[width=1\textwidth,height=0.4\textwidth]{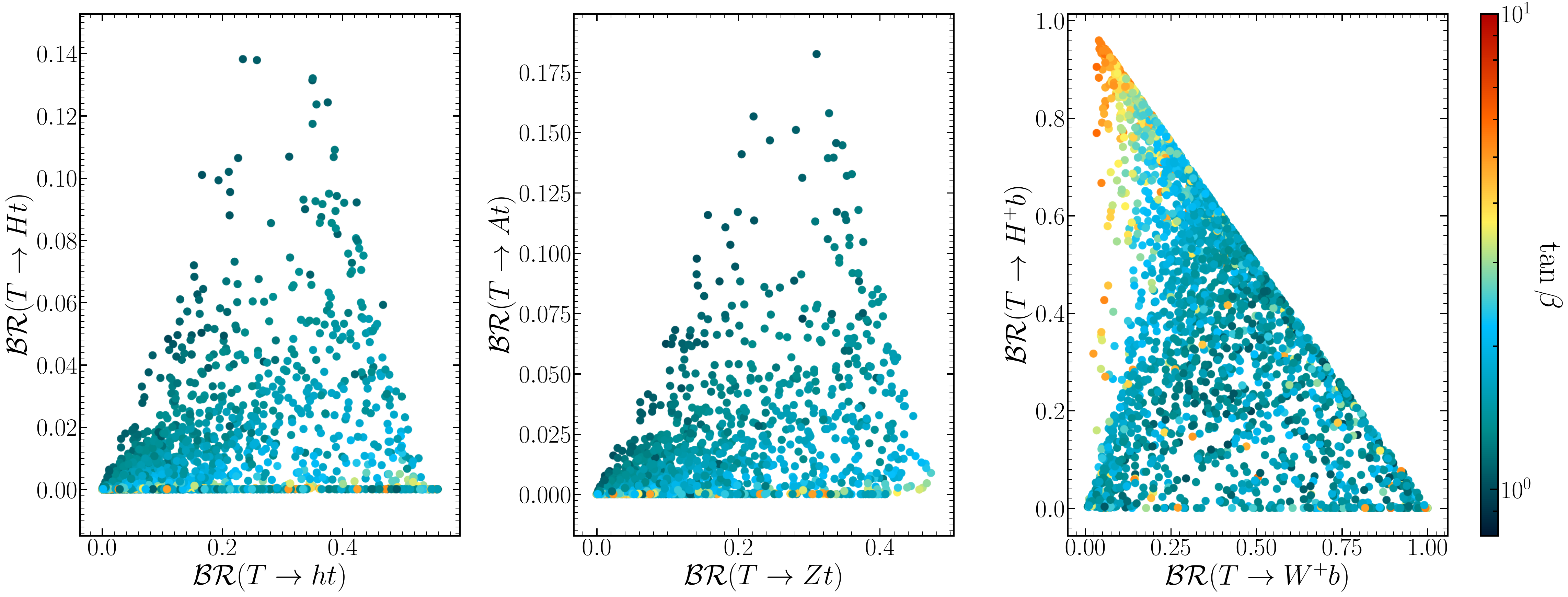} 	
	\caption{The correlation between ${\cal BR}(T\to ht)$ and ${\cal BR}(T\to Ht)$  (left),   
		${\cal BR}(T\to Zt)$ and ${\cal BR}(T\to At)$ (middle) as well as  ${\cal BR}(T\to W^+b)$ and ${\cal BR}(T\to H^+b)$ (right) with $\tan\beta$ indicated in the colour gauge.}	
	\label{fig7}
\end{figure*}
\begin{table*}[t!]
	\begin{center}
		\setlength{\tabcolsep}{30pt}
		\renewcommand{\arraystretch}{0.90}
		\begin{adjustbox}{max width=\textwidth}		
			\begin{tabular}{lccc}
				\toprule\toprule
				Parameters &       BP$_1$ &       BP$_2$ &       BP$_3$ \\\toprule
				\multicolumn{4}{c}{2HDM+VLQ inputs. The masses are in GeV.} \\\toprule

				$m_h$   &   125 &    125 &  125\\
				$m_H$  &   818.84 &  776.44 &  686.15\\
				$m_A$   &  622.86 &  593.82 &  621.9\\
				$m_{H\pm}$   &703.07 &  644.02 &  608.42  \\
				$\tan\beta$ &    1.37 &    1.59 &    2.15 \\
				$m_T$      & 965.31 & 1479.51 & 1863.45 \\
				$m_B$      & 969.75 & 1482.77 & 1872.40\\
				$\sin(\theta^u)_L$    & -0.0229 &    0.0119 &    0.0066  \\
				$\sin(\theta^d)_L$    &-0.0008 &   -0.0004 &   -0.0003 \\
				$\sin(\theta^u)_R$    &    -0.1273 &    0.1016 &    0.0712 \\
				$\sin(\theta^d)_R$    &   -0.1572 &   -0.1206 &   -0.1205\\
				\toprule
				\multicolumn{4}{c}{$\mathcal{BR}(H^\pm\to {XY})$ in \%} \\\toprule
				${\cal BR}(H^+\to t\bar{b})$ &99.73 & 99.80 & 99.73  \\
				${\cal BR}(H^+\to\tau\nu)$ &  0.02 &  0.03 &  0.10 \\
				${\cal BR}(H^+\to W^+ A)$ &  0.09 &  0.00 &  0.00 \\
				\toprule
				\multicolumn{4}{c}{$\mathcal{BR}(T\to {XY})$ in \%} \\\toprule
				${\cal BR}(T\to W^+b)$  &   47.92 & 29.30 & 19.860\\
				${\cal BR}(T\to W^+B)$  &   - &   - &   - \\
				${\cal BR}(T\to Zt)$  & 13.98 &  9.89 &  3.37 \\
				${\cal BR}(T\to ht)$  &17.22 & 10.84 &  3.58  \\
				${\cal BR}(T\to Ht)$  &0.00 &  0.91 &  0.13\\
				${\cal BR}(T\to At)$  &  1.10 &  1.08 &  0.12 \\
				${\cal BR}(T\to H^+b)$ & 19.78 & 47.98 & 72.94\\
				${\cal BR}(T\to H^+B)$ &   - &  - &   -  \\
				
				\toprule
				\multicolumn{4}{c}{Total decay width in GeV.} \\\toprule
				$\Gamma(T)$ &15.27 & 52.74 & 154.92\\\toprule
				\multicolumn{4}{c}{Observables} \\\toprule
				$T_{\mathrm{2HDM}}$  & -0.1644 &   -0.1177 &    0.0187  \\
				$T_{\mathrm{VLQ}}$  &   0.1605 &    0.1641 &    0.1123 \\
				$S_{\mathrm{2HDM}}$ &  0.0015 &    0.0034 &    0.0037\\
				$S_{\mathrm{VLQ}}$ & 0.0228 &    0.0169 &    0.0151 \\	
				$\Delta\chi^2(S_{\mathrm{2HDM+VLQ}},T_{\mathrm{2HDM+VLQ}})$ &    1.91 &    0.17 &    1.48\\\toprule
				$\chi^2{(h_{125})}\equiv\chi^2_\texttt{HiggsSignals}$ & 156.65 &  158.13 &  157.52 \\	\toprule\toprule
				
			\end{tabular}
		\end{adjustbox}
	\end{center}
	\caption{The full description of our BPs for the $(TB)$ doublet case.}\label{Bp3}
\end{table*}
In this representation of the 2HDM+VLQ, using the scan ranges in Tab.~\ref{table1}, we illustrate in 
Fig.~\ref{fig4} the $S$ and $T$ parameters and separate therein the VLQ contribution from the pure 2HDM one. From Fig.~\ref{fig4} (left), like in the previous case of singlet heavy top $(T)$, it is clear that the contribution of the 
VLQ $(TB)$  doublet and 2HDM states can be of opposite sign, which could then drastically modify the constraints on the parameter space of the model stemming from EWPOs. Further, from Fig.~\ref{fig4} (right), one can see that, for the SM with the VLQ contribution only, the mass splitting between $T$ and $B$ is required to be smaller than 5 GeV in the positive direction and 17 GeV in the negative one while in the case of the 2HDM with VLQs the splitting gets larger by almost a factor of 2. Therefore, we see again that the interplay between the additional VLQ and 2HDM states releases a substantial region of parameter space which would be otherwise unavailable to each BSM setup separately, crucially enabling VLQ decays into each other. Specifically, in the case of the SM with a VLQ $(TB)$ doublet, the new top can decay via one of the following modes: $T\to\{ W^+b , Zt,th,BW^{+}\}$. We remind the reader here that, for vanishing $\theta_R^t$, the couplings $tTZ$ and $tTh$  vanish and $T$  would decay dominantly into $W^+b$.  However, when the mixing does not vanish,  the above $T$ decays can proceed simultaneously with ${\cal BR}$'s dependent on the values of the mixing as well as of the VLQ masses. In the presence of extra Higgses, though, one can also have one or more of the following new channels open: $T\to\{At,Ht, H^+b, H^{+}B\}$  for $T$. We are keen to explore their relevance.

To illustrate the phenomenology of this BSM setup, we exploit again the systematic scan over the 2HDM parameters as well as 
on the VLQ $(TB)$ doublet ones described in Tab.~\ref{table1}, as usual, in the presence of all theoretical and experimental constraints, as previously discussed. Following such a scan, in Fig.~\ref{fig5}, we show the total width of the $T$ state as a function of the  $T$  mass (left frame) as well as of the up-quark sector mixing angle (right frame), both correlated to $\tan\beta$. In this setup, we see from the first plot that $\Gamma_T/m_T$ grows substantially with $\tan\beta$, so as to justify the beyond the Narrow Width Approximation (NWA) of Refs.~\cite{Carvalho:2018jkq,Moretti:2016gkr,Prager:2017hnt,Prager:2017owg,Moretti:2017qby},  it is worth noting from the second plot that this happens for rather large mixing $s^d_R$.

In Fig.~\ref{fig6}  we show the correlations between the $T$ decays into  SM   and non-SM particles, the latter involving one additional  Higgs state,  such as $T\to b H^\pm$, $T\to At$ and $T\to Ht$\footnote{In principle, in the definition of these two ${\cal BR}$'s, one should account for the $T\to W^+B$ and $T\to H^+B$ decays, however, because both the $W^+$ and $H^+$ are off-shell, they are always small, so we will not discuss these.}.  One can see from this figure that it is possible to have clear dominance for non-SM $T$ decays over a substantial region of parameter space.

Furthermore, in Fig.~\ref{fig7}, we look again at the correlations between the individual SM and non-SM decay channels, using the same pairing as previously. Herein, it is noticeable that, in the neutral current case, the latter is generally smaller than the former while, in the charged current case, $T\to H^+b$ can dominate $T\to W^+b$. One can also read that these last two decays are anti-correlated as a function of $\tan\beta$. The dominance of  $T\to W^+b$ corresponds to small, $\tan\beta$ while the dominance of $T\to H^+b$ prefers medium $\tan\beta$. 
In the left(middle)  panel of  Fig.~\ref{fig6}, it is clear that the maximum reach for ${\cal BR}(T\to ht)$ and ${\cal BR}(T\to Ht)$(${\cal BR}(T\to Zt)$ and ${\cal BR}(T\to At)$)  is around 60\% and 14\%(50\% and 18\%), respectively. In fact,  at large $\tan\beta$, the couplings $tTH$ and $tTA$ are suppressed, being proportional to $\cot\beta$. Therefore,
both $T\to At$ and $T\to Ht$ are generally suppressed in this limit, which makes $T\to Zt$ and $T\to ht$ rather substantial.

As done in the previous section, we finish this one by presenting again, in Tab.~\ref{Bp3}, three  BPs amenable to experimental investigation, wherein we target $T$ masses (with $\Gamma_T$ small)  just beyond the current experimental limit that we have found (BP$_3$) alongside two heavier mass values, in BP$_1$ and BP$_2$,  wherein one has a rather narrow and wide $T$ state, respectively, both of which might well be within the reach of the full  Run 3 of the LHC.

\subsection{2HDM with $(XT)$ doublet}
\begin{figure}[htpb!]
	\centering
	\includegraphics[width=0.45\textwidth,height=0.9\textwidth]{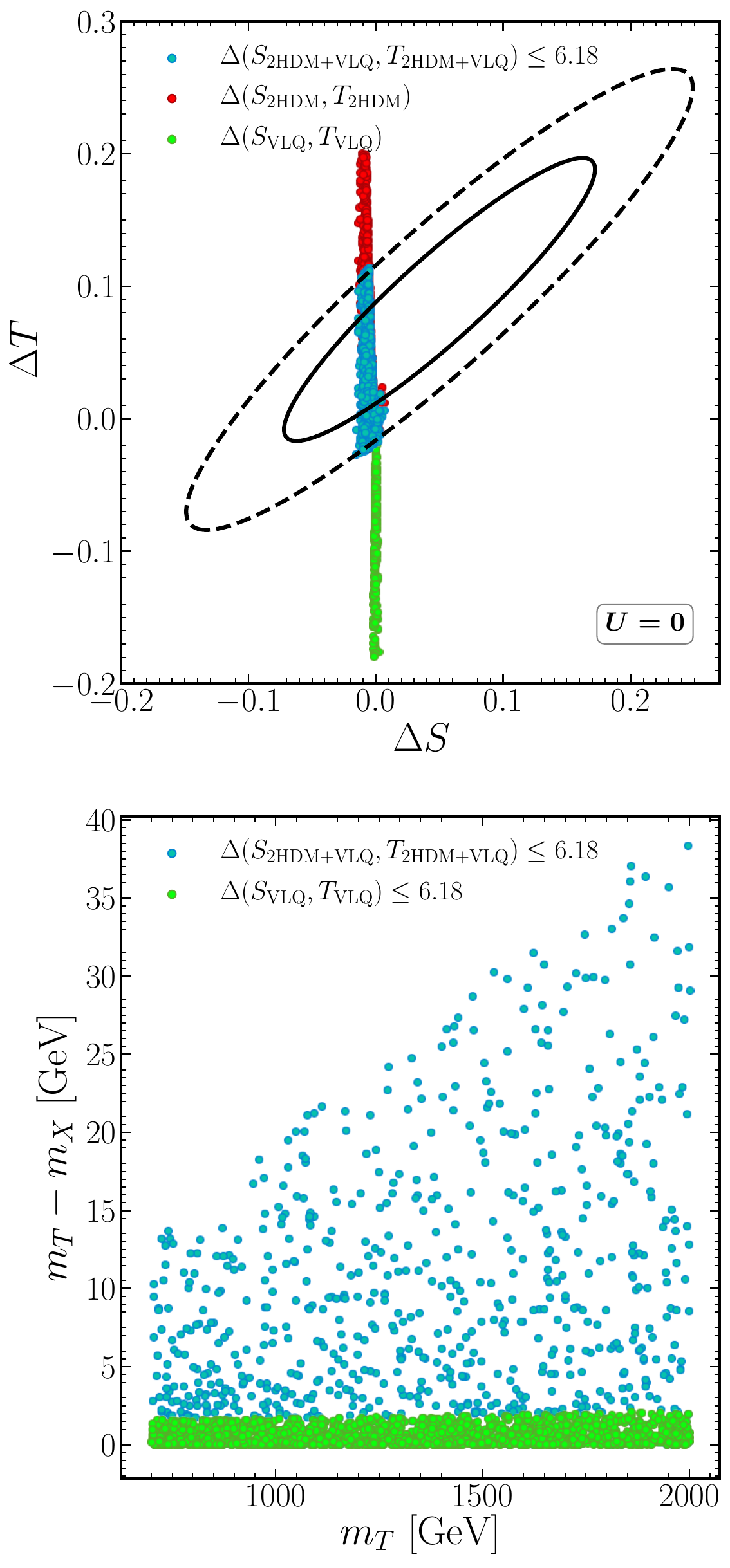}
	\caption{ (Top) Scatter plot of randomly generated points superimposed on the fit limits in the $(\Delta S, \Delta T)$ plane from EWPO data at the 95.45\% confidence level, with a correlation of 92\% \cite{ParticleDataGroup:2020ssz}. The black solid line marks the 1$\sigma$ CL region for $\chi_{ST}^2$, while the dashed line denotes the 2$\sigma$ CL region under the assumption $U=0$. Separate and combined contributions from the 2HDM and VLQ are displayed. 
		(Bottom) The same points are here mapped onto the $(m_T,\delta)$ plane, where $\delta$ is the mass difference between $T$ and $X$. Here, we only present the VLQ contribution and the total one. Further, all constraints have been taken into account.}
	\label{fig8}
\end{figure}
\begin{figure*}[htpb!]
	\centering 	
	\includegraphics[width=0.85\textwidth,height=0.45\textwidth]{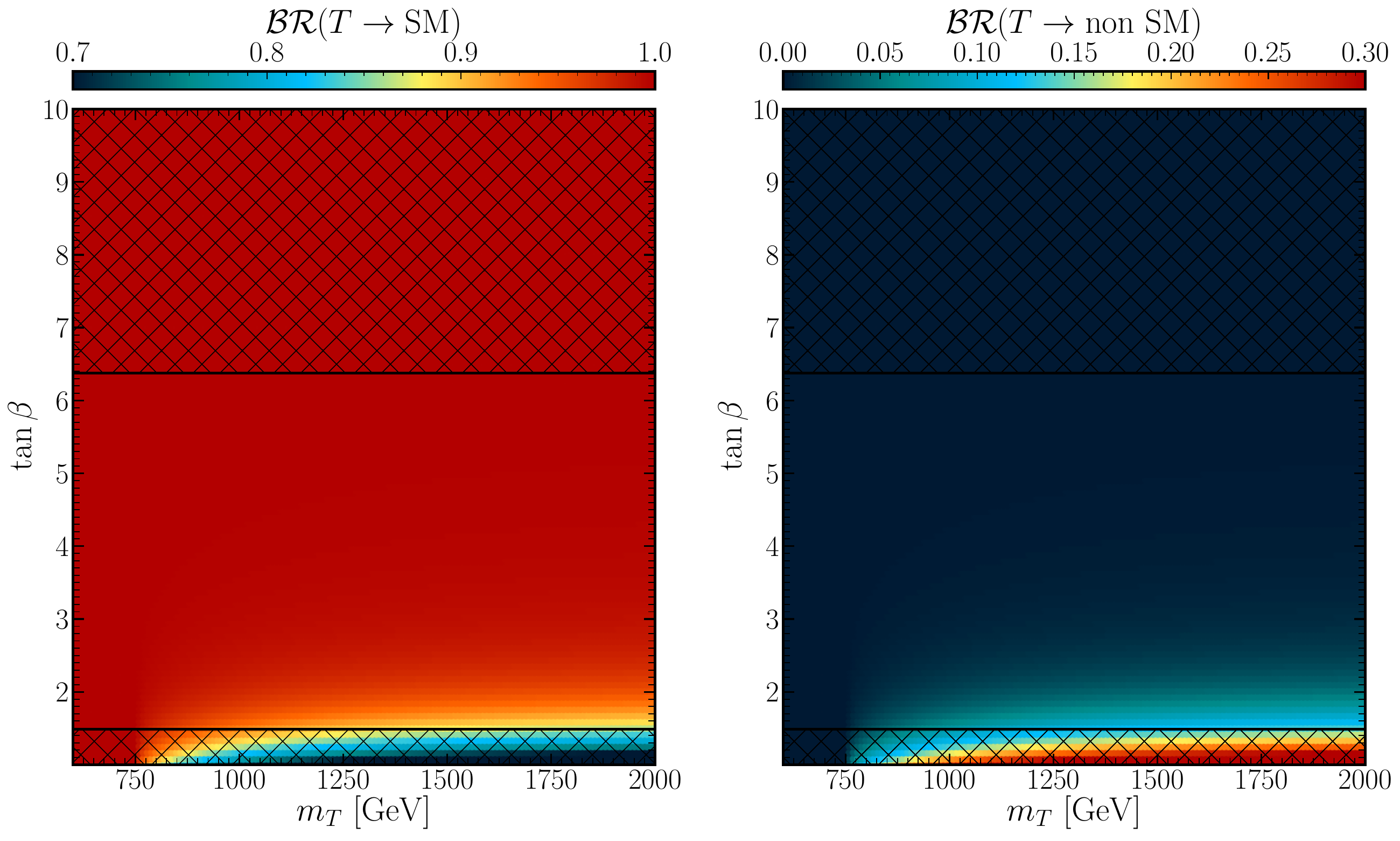}
	\caption{The ${\cal BR}(T\to$ SM) (left) and ${\cal BR}(T\to$ non SM)  (right) mapped onto the $(m_T, \tan\beta)$ plane. with $\sin\theta_R^u=0.057$ (the 2HDM parameters are the same as in Fig. \ref{fig3}). Here, the shaded areas are excluded by \texttt{HiggsBounds}, and all other constraints  ($S$, $T$, \texttt{HiggsSignals} and theoretical ones) are also checked.}	
	\label{fig9}
\end{figure*}
\begin{figure*}[htpb!]
	\centering
	\includegraphics[width=0.8\textwidth,height=0.40\textwidth]{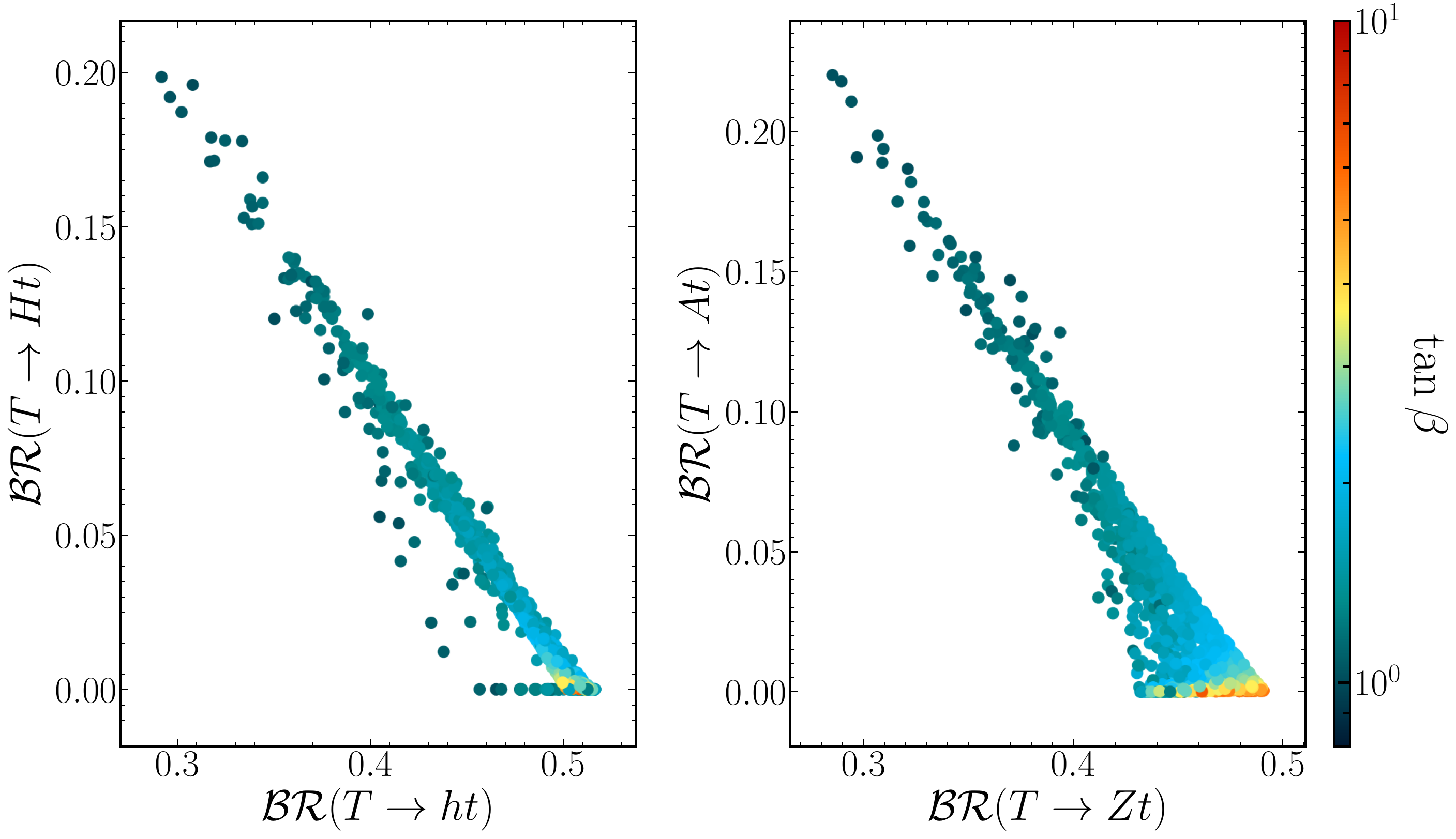}
	\caption{The correlation between ${\cal BR}(T\to ht)$ and ${\cal BR}(T\to Ht)$ (left),
		${\cal BR}(T\to Zt)$ and ${\cal BR}(T\to At)$ (right) with $\tan\beta$ as indicated in the colour gauge.}
	\label{fig10}
\end{figure*}
In the case of the SM extended with a $(XT)$ doublet, the ensuing VLQ structure is fully described by the $\theta_R^t$ mixing angle and the new top mass $m_T$.  In fact, for a  given $\theta_R^t$ value, $\theta_L^t$ is computed using 
Eq.~(\ref{ec:rel-angle1}). The mass of the new VLQ with exotic EM charge ($+5/3$), the $X$ state, is given as a function of such a mixing angle,  $m_T$, as well as $m_t$ by \cite{Aguilar-Saavedra:2013qpa}.

\begin{equation}
m_X^2=m_T^2\cos\theta_R+m_t^2\sin\theta_R.\label{eq_XT}
\end{equation}
This is independent (at tree level) from the additional parameters entering the 2HDM Higgs sector, however, the latter impinges on the viability of this BSM construct against EWPO data. 

Following the scan described in Tab.~\ref{table1}, in Fig.~\ref{fig8} (left), we demonstrate that even though the pure contributions of VLQ and 2HDM alone are (largely) out of the allowed 
EWPO data ellipses, the total contribution VLQ+2HDM falls within the allowed range for $\Delta S$ and $\Delta T$. In the right panel of Fig.~\ref{fig8}, we further illustrate the size of the mass splitting $\delta =m_T-m_X$ allowed by EWPO data. In the case of the VLQ-only structure, the splitting is always very small. Instead,  in the case of the full 2HDM+VLQ scenario, one can see that the splitting could be very large, of the order of 40 GeV, the more so the larger $m_T$. 

In Fig.~\ref{fig9}, we compare again ${\cal BR}(T\to {\rm SM})$ to ${\cal BR}(T\to {\rm non~SM})$, mapped over $m_T$ and $\tan\beta$. In contrast to the previous VLQ doublet realization, here, the latter are sub-leading concerning the former, altogether hardly relevant phenomenological and only very close to the edges of the available parameter space.

\begin{table*}[t!]
	\begin{center}
		\setlength{\tabcolsep}{45pt}
		\renewcommand{\arraystretch}{0.8}
		\begin{adjustbox}{max width=\textwidth}		
			\begin{tabular}{lcc}
				\toprule\toprule
				Parameters &       BP$_1$ &       BP$_2$ \\
				\toprule

				\multicolumn{3}{c}{2HDM+VLQ inputs. The masses are in GeV.} \\\toprule
				$m_h$   &   125&   125  \\
				$m_H$  &    836.11 &  850.43\\
				$m_A$   & 769.59 &  771.97 \\
				$m_{H\pm}$   &761.55 &  761.78  \\
				$\tan\beta$ &    1.10 &    1.08 \\
				
				$m_T$      &976.82 & 1879.55 \\
				
				$m_X$      &	761.55 &  761.78\\
				$\sin(\theta^u)_R$    &  -0.05 &   -0.04 \\
				
				\toprule
				\multicolumn{3}{c}{$\mathcal{BR}(H^\pm\to {XY})$ in \%} \\\toprule
				${\cal BR}(H^+\to t\bar{b})$     & 99.81 & 99.82 \\
				${\cal BR}(H^+\to\tau\nu)$ & 0.01 &  0.01 \\
				${\cal BR}(H^+\to W^+ A)$ &  0.01 &  0.00\\
				\toprule
				\multicolumn{3}{c}{$\mathcal{BR}(T\to {XY})$ in \%} \\\toprule
				${\cal BR}(T\to W^+b)$  &   2.80 &  0.56 \\
				${\cal BR}(T\to W^+B)$  &   - &   - \\
				${\cal BR}(T\to Zt)$  &  45.48 & 33.29 \\
				${\cal BR}(T\to ht)$  &   49.85 & 34.13   \\
				${\cal BR}(T\to Ht)$  &    0.00 & 15.94   \\
				${\cal BR}(T\to At)$  &   1.87 & 16.08 \\
				${\cal BR}(T\to H^+b)$ &  - &  - \\
				${\cal BR}(T\to H^+B)$ &   - &  -   \\				
				\toprule
				\multicolumn{3}{c}{Total decay width in GeV.} \\\toprule
				$\Gamma(T)$ & 0.72 &  5.96\\
				\toprule
				\multicolumn{3}{c}{Observables} \\\toprule
				$T_{\mathrm{2HDM}}$  &  0.0107 &    0.0161 \\
				$T_{\mathrm{VLQ}}$  &  -0.0212 &   -0.0290 \\
				$S_{\mathrm{2HDM}}$ &  0.0027 &    0.0032\\
				$S_{\mathrm{VLQ}}$ &   0.0000 &   -0.0004   \\	
				$\Delta\chi^2(S_{\mathrm{2HDM+VLQ}},T_{\mathrm{2HDM+VLQ}})$ &  5.53 &  5.95\\\toprule
				$\chi^2{(h_{125})}\equiv \chi^2_{\texttt{HiggsSignals}}$ &  158.69 &  158.66  \\	
				
				\bottomrule\bottomrule
				
			\end{tabular}
		\end{adjustbox}
	\end{center}
	\caption{The full description of our BPs for the $(XT)$ doublet case.}\label{Bp4}
\end{table*}

We now discuss the size of the individual ${\cal BR}$'s of $T$ decays. As usual, alongside the SM decays of new top $T$, into $\{W^+b, Zt, ht\}$, one has the non-SM decays $T\to Ht$ and $T\to A t$: remarkably, in fact, the $t\to H^+b$ channel is not available, as the intervening coupling is identically zero\footnote{Since ${\cal BR}(T \to H^+b)$ is zero, we do not display it in the figures.}.
In Fig.~\ref{fig10}, we illustrate the correlations between ${\cal BR}(T \to Zt)$ and ${\cal BR}(T \to At)$ (right), ${\cal BR}(T \to ht)$, and ${\cal BR}(T \to Ht)$ (left).  At large $\tan\beta$, both ${\cal BR}(T\to Ht)$ and ${\cal BR}(T\to At)$ are suppressed, leading to both ${\cal BR}(T\to Zt)$ and ${\cal BR}(T\to ht)$ reaching their SM values.
At small $\tan\beta$, instead, one can see that both ${\cal BR}(T\to Ht)$ and ${\cal BR}(T\to At)$ 
are somewhat enhanced and therefore ${\cal BR}(T\to Zt)$ and ${\cal BR}(T\to ht)$ are somewhat suppressed, highlighting this region of parameter space as being the most suitable one for exotic $T$ decay searches (albeit limited to neutral decay currents in this BSM framework). However, in line with the previous figure, none of the exotic (neutral) decay channels is ever very large, never passing the 25\% or so ${\cal BR}$ value.

Finally, the BPs that we recommend for further phenomenological investigation of this BSM scenario are found in Tab.~\ref{Bp4}, including both a light (BP$_1$) and heavy (BP$_2$) $T$ state.

\subsection{2HDM with $(XTB)$ triplet}

We discuss here the $(XTB)$ triplet case.  Before presenting our numerical results, though,  let us first introduce our parametrization.
This model is fixed by giving the new top mass and one mixing angle, let us say $\theta_L^t$, the other parameters are then computable. In fact, $\theta_R^t$ is 
derived from Eq. (\ref{ec:rel-angle1}) while $m_X$ is given 
by \cite{Aguilar-Saavedra:2013qpa}:

\begin{eqnarray}
m_X^2=m_T^2\cos\theta^u_L+m_t^2 \sin\theta^u_L = m_B ^2 \cos^2\theta_L^b+m_b^2 \sin^2\theta_L^b.
\end{eqnarray}

Using the above relation between $m_T$ and $m_X$ together with the one between up- and down-type quark mixing in Eq.~(\ref{TBY-mix}), one can derive the mass
of the new bottom quark as follows:
\begin{eqnarray}
m_B^2= \frac{1}{2}\sin^2(2\theta^u_L)(m_T^2-m_t^2)^2/(m_X^2-m_b^2)+m_X^2.
\end{eqnarray}
The down-type quark mixing is then given by.
\begin{eqnarray}
\sin(2\theta_L^d) = 
\sqrt{2}\frac{m_T^2-m_t^2}{m_B^2-m_b^2}\sin (2\theta^u_L).
\end{eqnarray}

\begin{figure*}[htpb!]
	\centering
	\includegraphics[width=0.45\textwidth,height=0.45\textwidth]{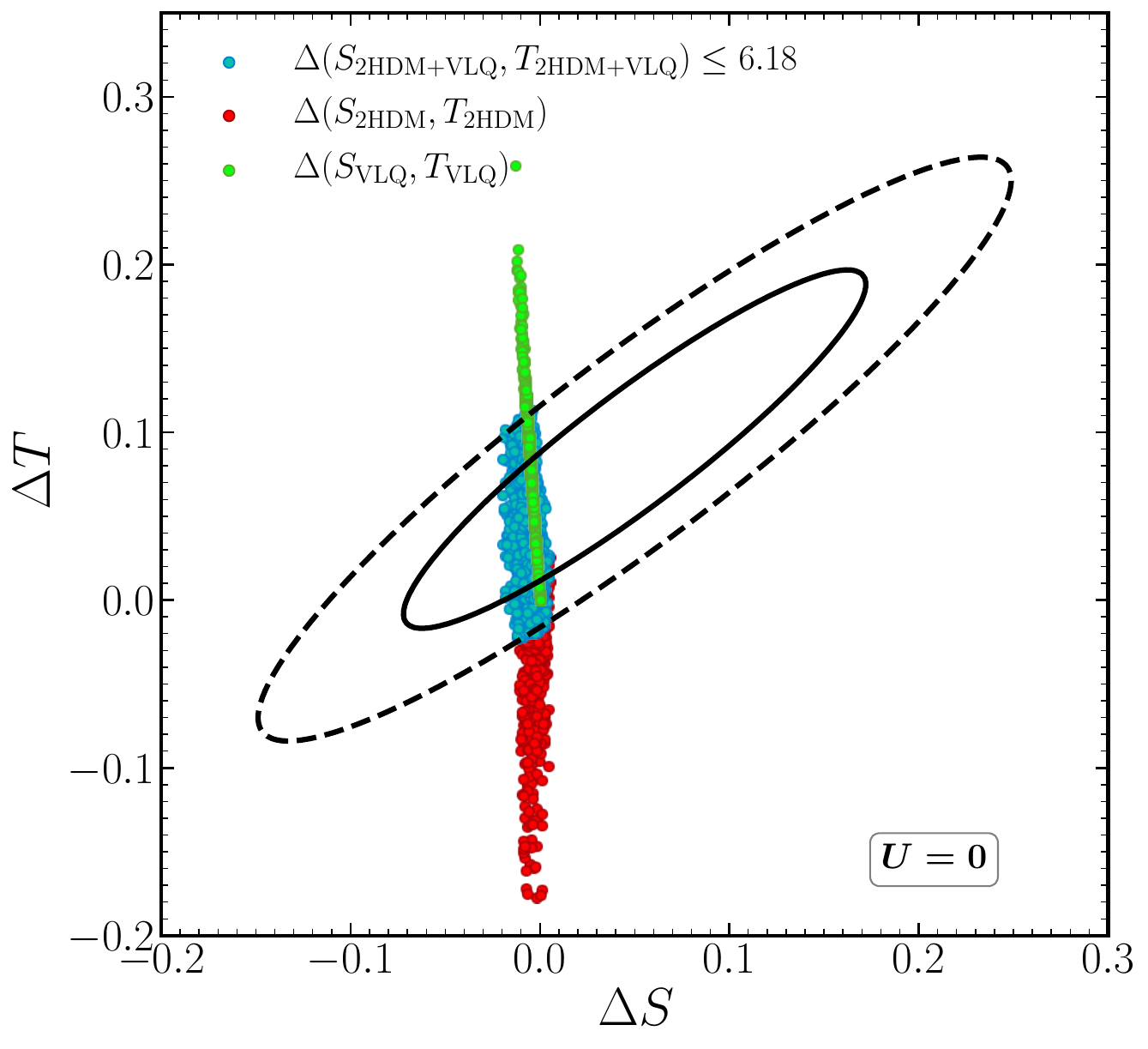}
	\caption{ Scatter plot of randomly generated points superimposed on the fit limits in the $(\Delta S, \Delta T)$ plane from EWPO data at the 95.45\% confidence level, with a correlation of 92\% \cite{ParticleDataGroup:2020ssz}. The black solid line marks the 1$\sigma$ CL region for $\chi_{ST}^2$, while the dashed line denotes the 2$\sigma$ CL region under the assumption $U=0$. Separate and combined contributions from the 2HDM and VLQ are displayed.  Further, all constraints have been taken into account.}
	\label{fig11}
\end{figure*}

As usual, we perform a systematic scan over both the 2HDM and VLQ parameters, as indicated in
Tab.~\ref{table1}. In Fig.~\ref{fig11}, we show that, even if 2HDM and VLQ points (mostly)
fall out of the allowed  $\Delta S$ and $\Delta T$ ellipses when plotted separately, after adding the 2HDM and VLQ contributions together, one observes that the total contribution can well be within the allowed range. 

\begin{figure*}[htpb!]
	\centering
	\includegraphics[width=0.85\textwidth,height=0.45\textwidth]{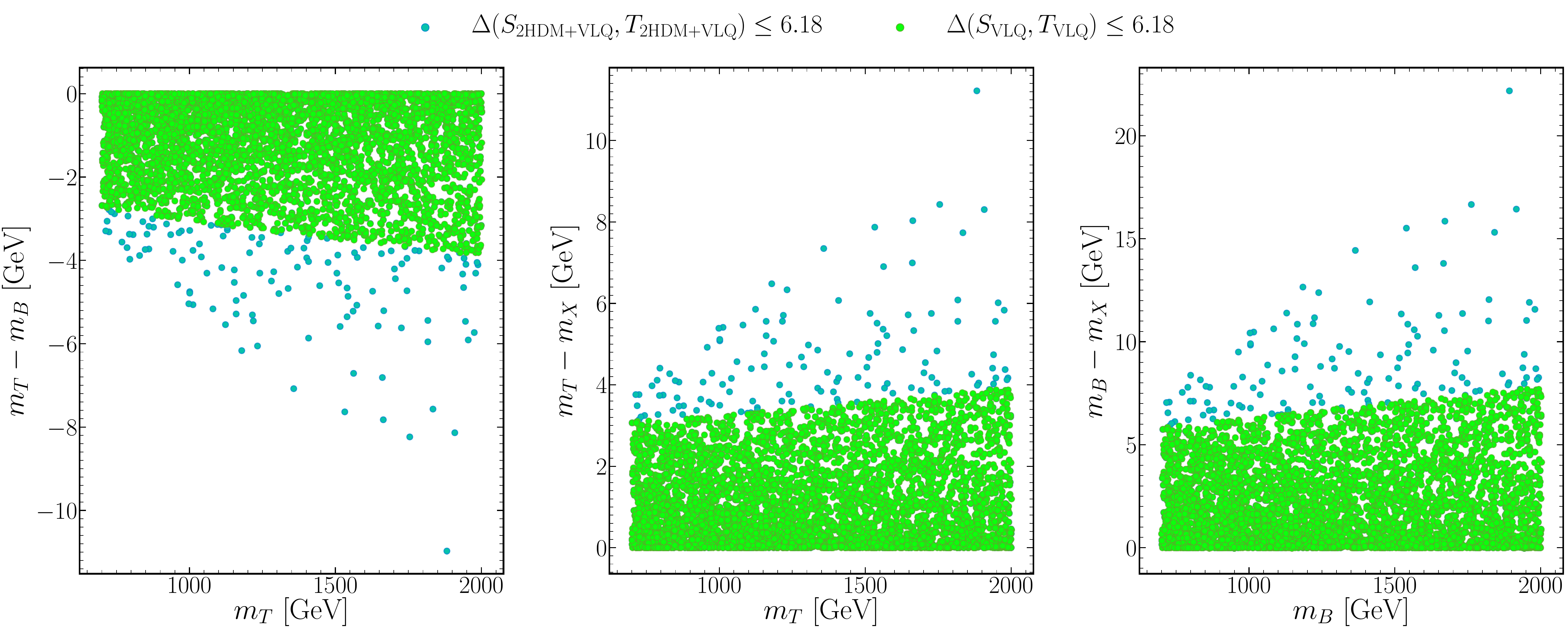}
	\caption{Scatter plots of randomly generated points mapped onto the $(m_{T/B},\delta)$ plane, where $\delta$ is the mass difference between $T$ and $B$, $T$ and $X$, and $B$ and $X$.  Here,  we illustrate the VLQ and 2HDM+VLQ contributions separately. Further, all constraints have been taken into account.}
	\label{fig12}
\end{figure*}

Furthermore, the splitting between $m_T,m_B$, $m_T,m_X$ and $m_B,m_X$ is severely constrained by the EWPO constraints (this can be seen in Fig.~\ref{fig12}, respectively, mapped against $m_T$), one can notice that such a splitting is rather small in the case of the SM  with VLQ construct,  of order a few GeV. However, in the case of the full 2HDM with VLQ scenario, such a splitting becomes large at most 11 or 22 GeV, thereby signalling that inter-VLQ decays are bound to play a negligible role in this BSM realization.  

\begin{figure*}[htpb!]
	\centering
	\includegraphics[width=0.85\textwidth,height=0.45\textwidth]{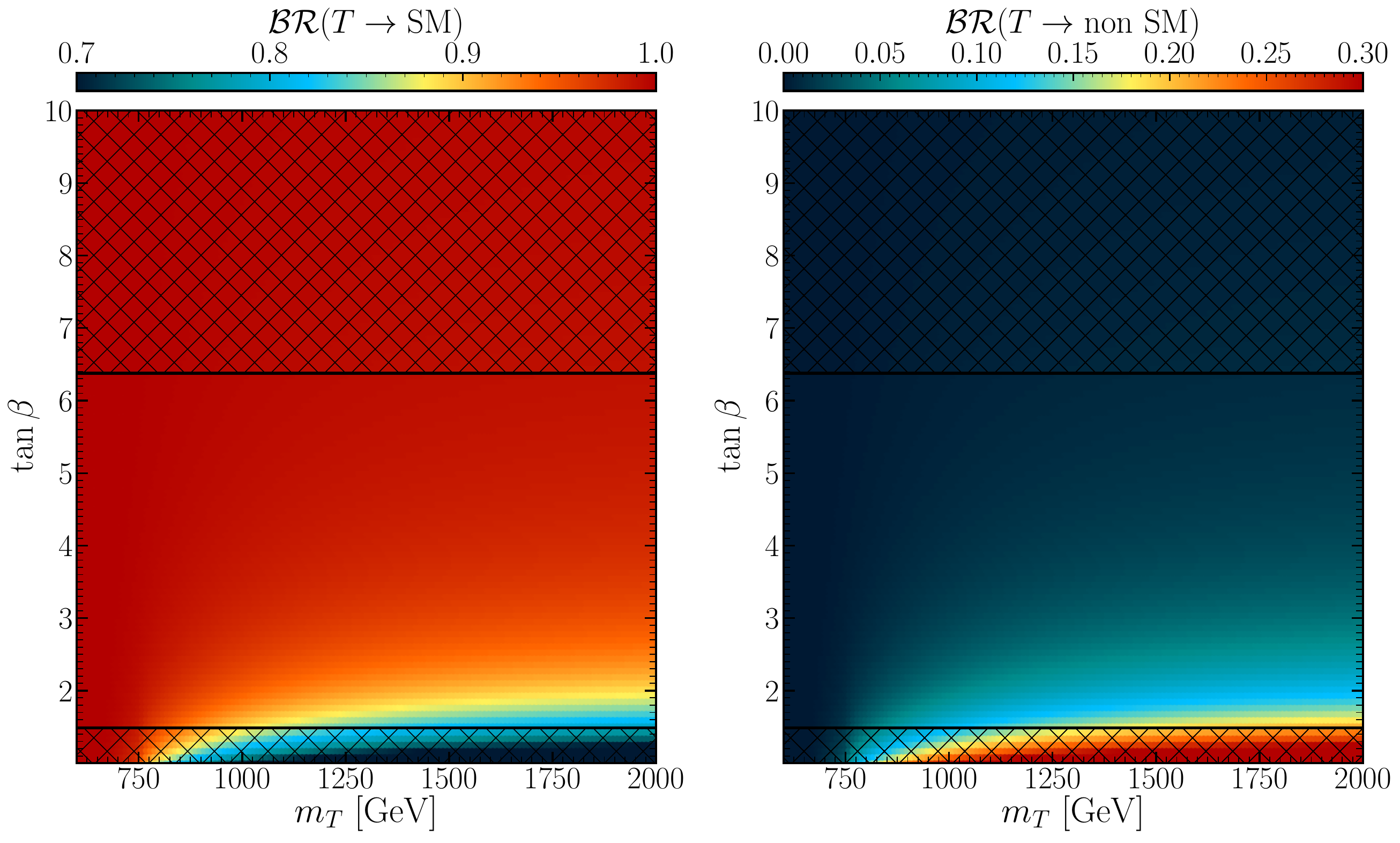} 
	\caption{The ${\cal BR}(T\to$ SM) (left) and ${\cal BR}(T\to$ non SM)  (right) mapped onto the $(m_T, \tan\beta)$ plane, with $\sin\theta_L^u=0.0093$ (the 2HDM parameters are the same as in Fig. \ref{fig3}). Here, the shaded areas are excluded by \texttt{HiggsBounds}, and all other constraints  ($S$, $T$, \texttt{HiggsSignals} and theoretical ones) are also checked.}	
	\label{fig13}
\end{figure*}

In the case of the 2HDM with a $(XTB)$ triplet, the new top can decay into the following SM channels: $\{Zt, ht, W^+b\}$. Further, it can do so also via the extra channels $\{Ht, At, H^+b\}$, which involves the additional Higgs states of the 2HDM. The relative importance of the former concerning the latter 
is illustrated in Fig.~\ref{fig13}, wherein the dominance of the SM channels is manifest, although the non-SM ones can reach the 20\% level in the cumulative
${\cal BR}$'s. We now discuss these individual $T$ decay channels. (Note that, because of the small splitting between $X$ and $T$,  the decay 
$T\to W^+ X$ is closed for a real $W^+$ and is very suppressed for an off-shell one, so we do not discuss it here.)

\begin{figure*}[h!]
	\centering
	\includegraphics[width=1\textwidth,height=0.40\textwidth]{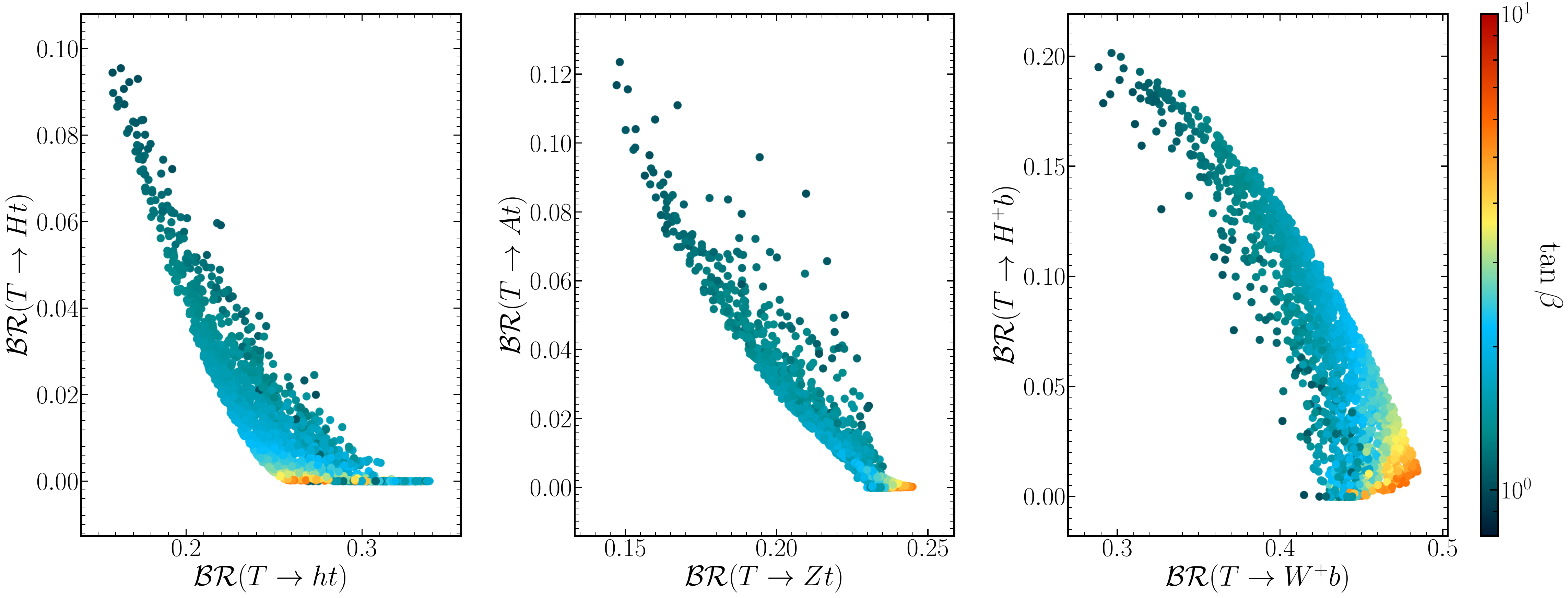}
	\caption{The correlation between ${\cal BR}(T\to ht)$ and ${\cal BR}(T\to Ht)$ (left),   
		${\cal BR}(T\to Zt)$ and ${\cal BR}(T\to At)$ (middle) as well as   ${\cal BR}(T\to W^+b)$ and ${\cal BR}(T\to H^+b)$ (right) with $\tan\beta$ indicated in the colour gauge.}	
	\label{fig14}
\end{figure*}

As intimated, the presence of the extra channels, specific to the additional 2HDM states, can change somewhat the SM picture where the new top can decay into one of these final states: $\{Zt, ht, W^+b\}$. In the full 2HDM+VLQ, the new channels $T\to \{ Ht, At, H^+b\}$ could become significant, which implies that simultaneously the SM decays of the new top become suppressed. In fact, 
we stress that the $HTt$, $ATt$ and $hTt$ couplings have a term which 
is proportional to $\cot\beta \sin(\beta-\alpha)$ and this implies that, near the decoupling limit, i.e., $\sin(\beta-\alpha)\approx 1$, those couplings become significant for small $\tan\beta$. Note that the left-handed component of the $H^+Tb$ coupling also has this $\cot\beta$ factor and would enjoy the same enhancement for small $\tan\beta$. These patterns are illustrated in Fig.~\ref{fig14}, which however makes it clear that the role of the exotic decays is only relevant at the level of 20\% at the most (as expected from the previous figure).

Again, we present our  BPs in a table, Tab.~\ref{Bp6}, two in particular, one with a light $T$ state (BP$_2$) and one with a heavy $T$ state (BP$_1$), of varying width,  in line with the previous subsections.

\begin{table*}[t!]
	\begin{center}
		\setlength{\tabcolsep}{45pt}
		\renewcommand{\arraystretch}{0.8}
		\begin{adjustbox}{max width=\textwidth}		
			\begin{tabular}{lcc}
				\toprule\toprule
				Parameters &       BP$_1$ &       BP$_2$ \\
				\toprule
				\multicolumn{3}{c}{2HDM+VLQ inputs. The masses are in GeV.} \\\toprule
				$m_h$   &   125&   125  \\
				$m_H$  &   927.96 & 780.97 \\
				$m_A$   &  725.58 & 531.95 \\
				$m_{H\pm}$   &  834.63 & 696.72 \\
				$\tan\beta$ &    1.01 &   1.43 \\
				
				$m_T$      &1576.63 & 987.10 \\
				$m_B$      & 1584.20 & 993.45  \\
				$m_X$      & 1568.83 & 980.28  \\
				$\sin(\theta^u)_L$    &   -0.10 &   0.12 \\
				$\sin(\theta^d)_L$    &  -0.14 &   0.16 \\
				\toprule
				\multicolumn{3}{c}{$\mathcal{BR}(H^\pm\to {XY})$ in \%} \\\toprule
				${\cal BR}(H^+\to t\bar{b})$     & 97.53 & 73.66 \\
				${\cal BR}(H^+\to\tau\nu)$ & - &  - \\
				${\cal BR}(H^+\to W^+ A)$ &  2.30 & 26.20\\
				\toprule
				\multicolumn{3}{c}{$\mathcal{BR}(T\to {XY})$ in \%} \\\toprule
				${\cal BR}(T\to W^+b)$  &   31.92 & 42.01 \\
				${\cal BR}(T\to W^+B)$  &   - &   - \\
				${\cal BR}(T\to Zt)$  &    16.32 & 22.02 \\
				${\cal BR}(T\to ht)$  &17.70 & 26.89 \\
				${\cal BR}(T\to Ht)$  & 7.31 &  0.63  \\
				${\cal BR}(T\to At)$  &   9.46 &  2.41 \\
				${\cal BR}(T\to H^+b)$ & 17.29 &  6.03 \\
				${\cal BR}(T\to H^+B)$ &   - &  -   \\				
				\toprule
				\multicolumn{3}{c}{Total decay width in GeV.} \\\toprule
				$\Gamma(T)$ & 37.56 &  9.14 \\
				\toprule
				\multicolumn{3}{c}{Observables} \\\toprule
				$T_{\mathrm{2HDM}}$  &  -0.1804 &  -0.2443  \\
				$T_{\mathrm{VLQ}}$  &  0.2723 &   0.2579 \\
				$S_{\mathrm{2HDM}}$ & -0.0003 &  -0.0026 \\
				$S_{\mathrm{VLQ}}$ &   -0.0143 &  -0.0154  \\	
				$\Delta\chi^2(S_{\mathrm{2HDM+VLQ}},T_{\mathrm{2HDM+VLQ}})$ &   4.5133 &   1.3442\\\toprule
				$\chi^2{(h_{125})}\equiv \chi^2_{\texttt{HiggsSignals}}$ &  158.35 & 158.29 \\	
				
				\bottomrule\bottomrule
				
			\end{tabular}
		\end{adjustbox}
	\end{center}
	\caption{The full description of our BPs for the $(XTB)$ triplet case.}\label{Bp6}
\end{table*}

\subsection{2HDM with $(TBY)$ triplet}
We finally discuss the $(TBY)$ triplet case. In the 2HDM with such a VLQ representation,  the situation is very similar to the case of the ($T$) singlet and $(TB)$ doublet. Before discussing our numerical results, though,  let us again first introduce our parametrization.
This model is fixed by giving the new top mass and one mixing angle, let us say $\theta_L^t$, the other parameters are then computable. In fact, $\theta_R^t$ is 
derived from Eq. (\ref{ec:rel-angle1}) while $m_Y$ is given 
by \cite{Aguilar-Saavedra:2013qpa}:
\begin{eqnarray}
m_Y^2 &=& m_T^2 \cos^2\theta_L^t+m_t^2 \sin^2\theta_L^t  = 
m_B ^2 \cos^2\theta_L^b+m_b^2 \sin^2\theta_L^b.
\end{eqnarray}
Using the above relation between $m_T$ and $m_Y$ together with the one between up- and down-type quark mixing in Eq.~(\ref{TBY-mix}), one can derive the mass
of the new bottom quark as:
\begin{eqnarray}
m_B^2 &=& \frac{1}{8} \sin^2 2 \theta_L^t \frac{(m_T^2 -m_t^2)^2}{m_Y^2-m_b^2}+m_Y^2.
\end{eqnarray}
With this in hand,  one can then derive the down-type quark mixing $\theta_{L,R}^d$ 
using Eqs.~(\ref{ec:rel-angle1})--(\ref{TBY-mix}).

\begin{figure*}[htpb!]
	\centering
	\includegraphics[width=0.45\textwidth,height=0.45\textwidth]{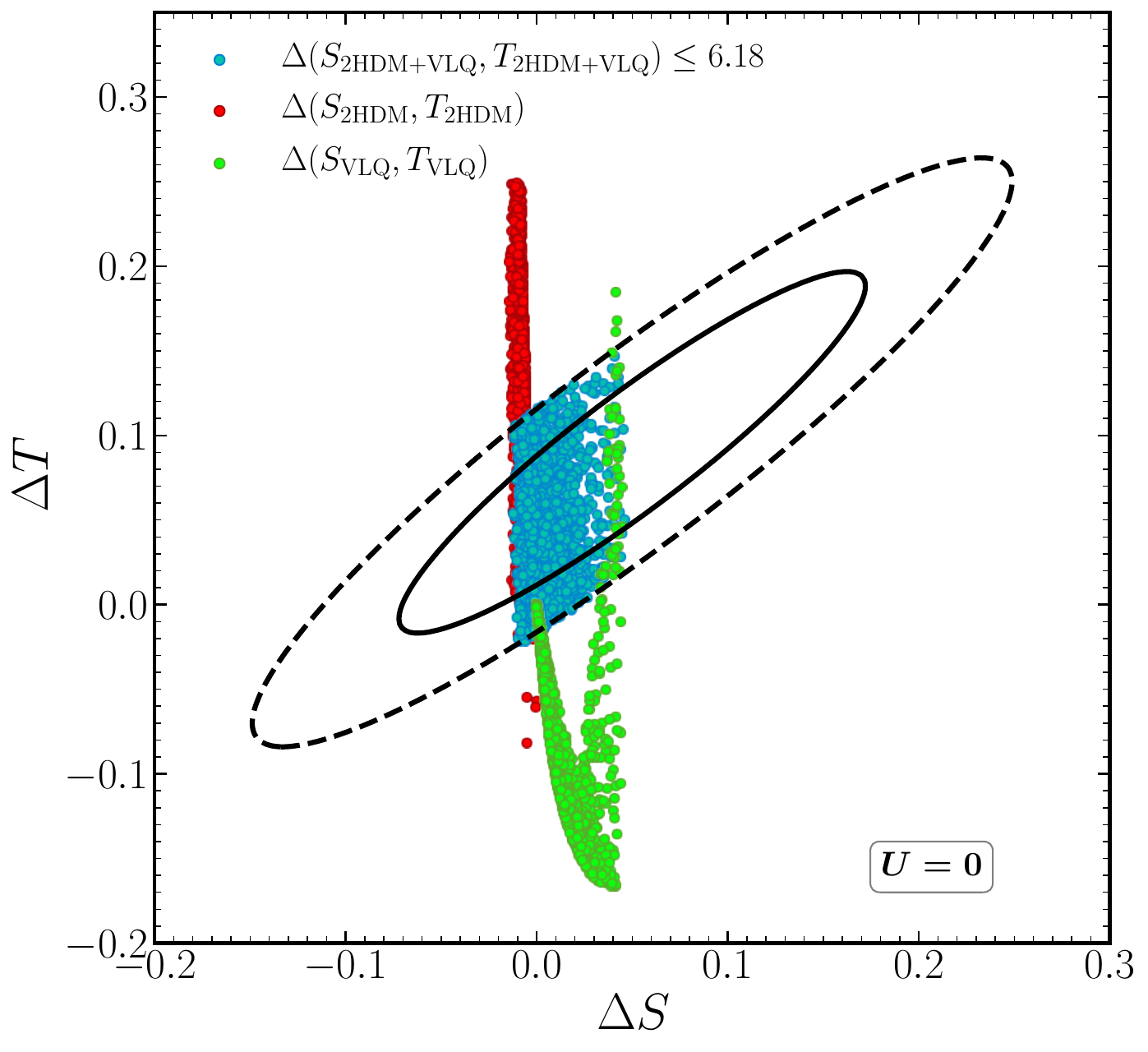}
	\caption{ Scatter plot of randomly generated points superimposed on the fit limits in the $(\Delta S, \Delta T)$ plane from EWPO data at the 95.45\% confidence level, with a correlation of 92\% \cite{ParticleDataGroup:2020ssz}. The black solid line marks the 1$\sigma$ CL region for $\chi_{ST}^2$, while the dashed line denotes the 2$\sigma$ CL region under the assumption $U=0$. Separate and combined contributions from the 2HDM and VLQ are displayed.  Further, all constraints have been taken into account.}
	\label{fig15}
\end{figure*}

Based on the scan listed in  Tab.~\ref{table1}, we   
first illustrate in Fig.~\ref{fig15}   the allowed 95\% CL regions  from the  
$S$ and $T$ parameter constraints, which shows that both the 2HDM only and VLQ only  
contributions could well be out of the allowed ranges but, when adding these together, one indeed finds viable solutions because of cancellations.
As mentioned previously, when adding VLQs alongside extra Higgs states,  the phenomenology of the $S$ and $T$ can change drastically, with respect to the SM case. 
\begin{figure*}[htpb!]
	\centering
	\includegraphics[width=0.85\textwidth,height=0.4\textwidth]{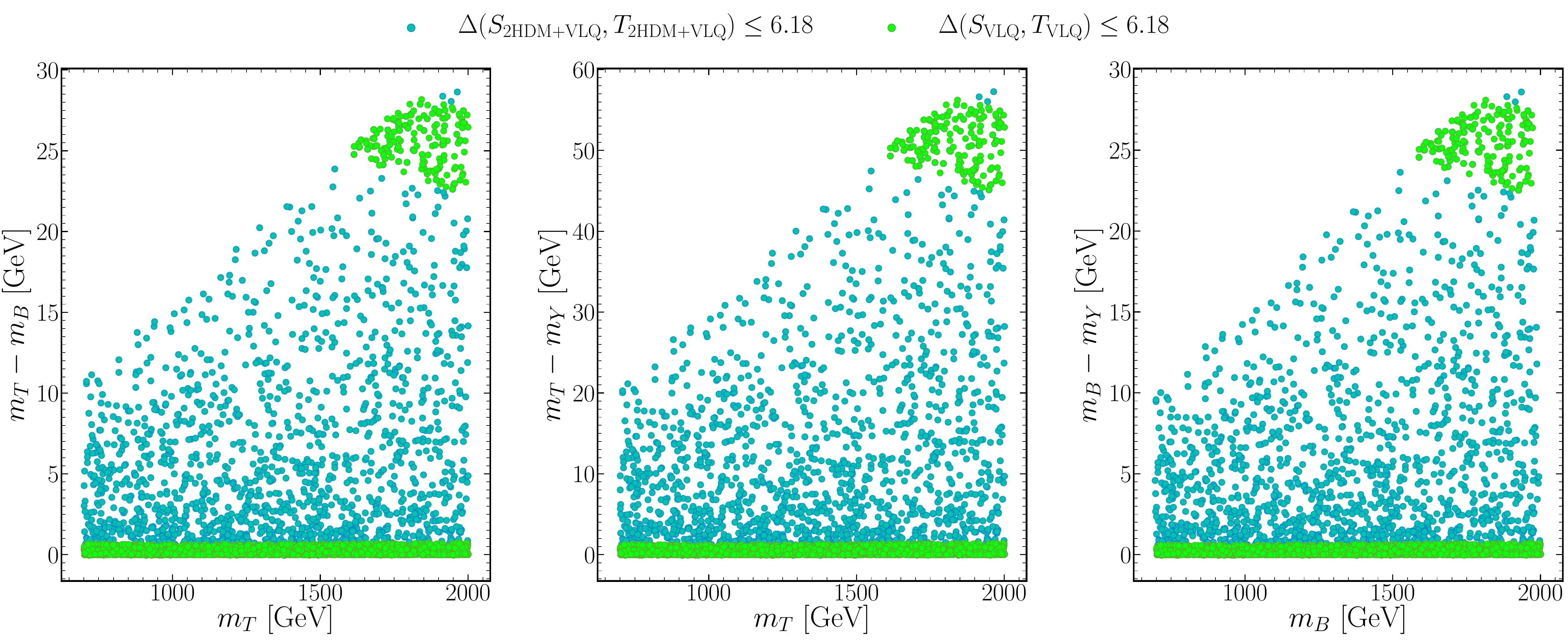}
	\caption{Scatter plots of randomly generated points mapped onto the $(m_{T/B},\delta)$ plane, where $\delta$ is the mass difference between $T$ and $B$, $T$ and $Y$, and $B$ and $Y$.  Here,  we illustrate the VLQ and 2HDM+VLQ contributions separately. Further, all constraints have been taken into account.}
	\label{fig16}
\end{figure*}

\begin{figure*}[htpb!]
	\centering
	\includegraphics[width=0.85\textwidth,height=0.45\textwidth]{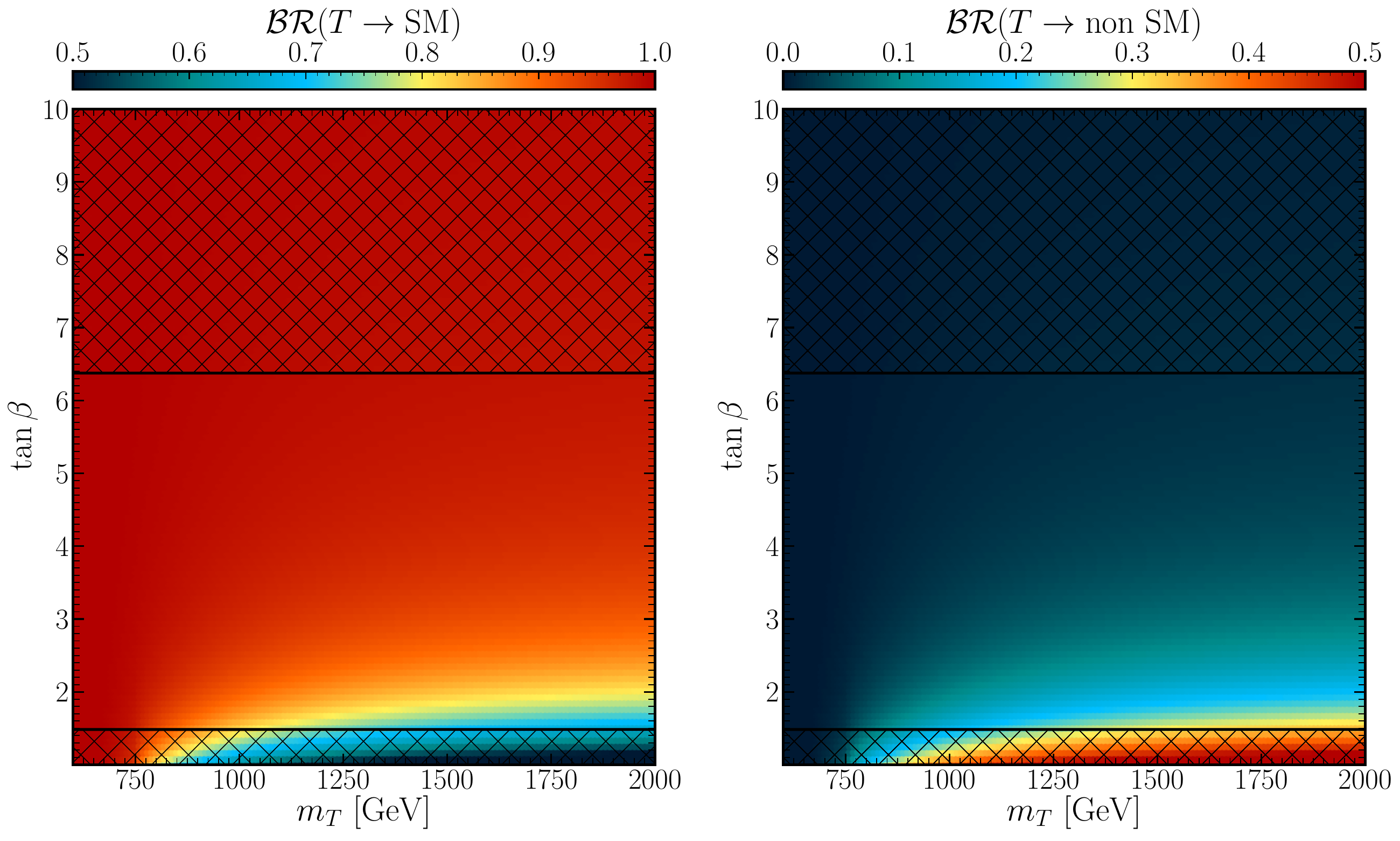} 
	\caption{The ${\cal BR}(T\to$ SM) (left) and ${\cal BR}(T\to$ non SM)  (right) mapped onto the $(m_T, \tan\beta)$ plane, with $\sin\theta_L^u=0.02$ (the 2HDM parameters are the same as in Fig. \ref{fig3}). Here, the shaded areas are excluded by \texttt{HiggsBounds}, and all other constraints  ($S$, $T$, \texttt{HiggsSignals} and theoretical ones) are also checked.}	
	\label{fig17}
\end{figure*}
\begin{figure*}[htpb!]
	\centering
	\includegraphics[width=1\textwidth,height=0.40\textwidth]{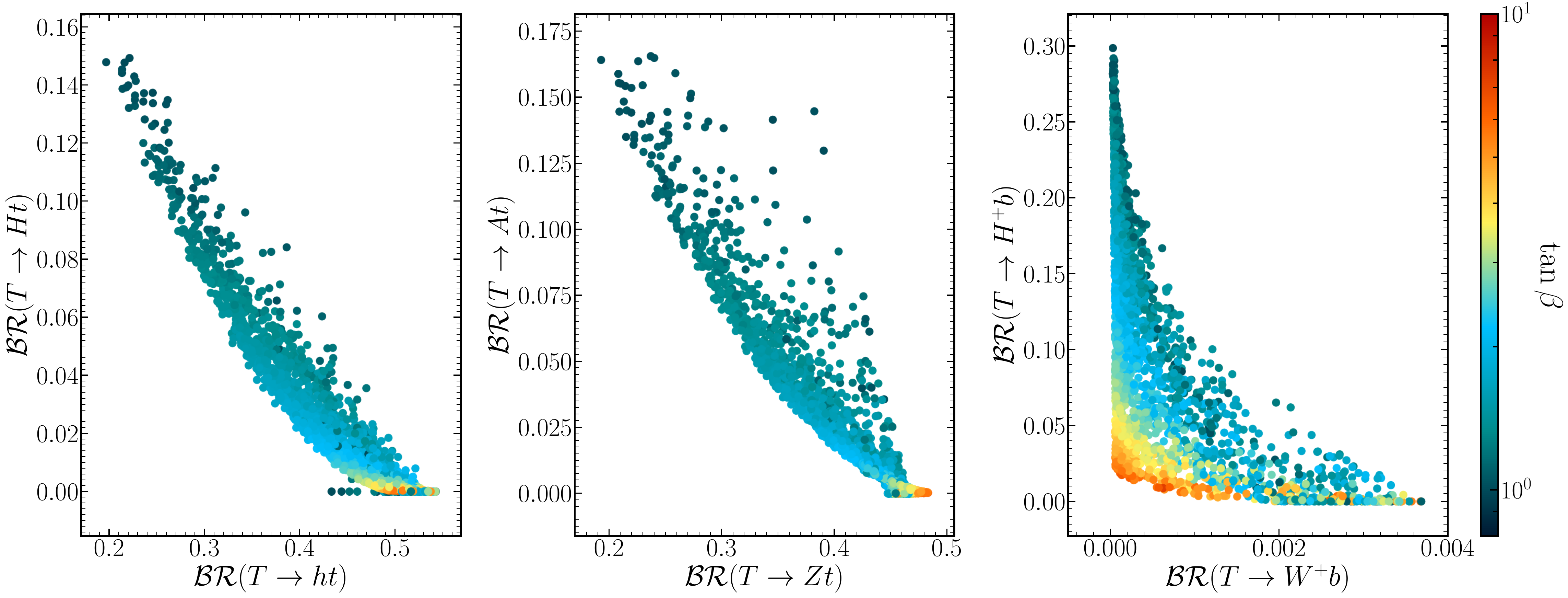}
	\caption{The correlation between  ${\cal BR}(T\to ht)$ and ${\cal BR}(T\to Ht)$ (left)  (right),   
		${\cal BR}(T\to Zt)$ and ${\cal BR}(T\to At)$ (middle) as well as  ${\cal BR}(B\to W^-t)$ and ${\cal BR}(B\to H^-t)$(left) with $\tan\beta$ indicated in the colour gauge.}	
	\label{fig18}
\end{figure*}
Following Fig.~\ref{fig16}, in the SM with this considered VLQ representation, the splitting $\delta$ between the masses of $T$ and $B$ (left), 
$T$ and $Y$ (middle) as well as $B$ and $Y$ (right) is rather small,  of the order a fraction of GeV at small $m_T$. However, for high $m_T$ values, these splittings  become more pronounced, reaching up to 28.2 GeV between $T$-$B$ and $B$-$Y$, and as much as 56 GeV for $T$-$Y$. Similarly, in the 2HDM+VLQ case, the situation exhibits comparable consistency. The mass splitting between $T$ and $Y$ can exceed 57 GeV, while the splittings between $T$-$B$ and $B$-$Y$ can go up to 28.6 GeV. Altogether, yet again, the substantial cancellations between additional Higgs and VLQ states, naturally occurring in loop contributions owing to their different spin statistics, enable one to gain significant parameter space, that we will therefore, in line with the
previous two subsections, exploit to study 2HDM decays of VLQ states. Again, though, the small $\delta$ values seen in these plots exemplify the fact the VLQ decays into each other are, yet again, mostly negligible.

In the SM with a $(TBY)$ triplet, the decay patterns of $T$  states are essentially the same as in the $(TB)$ doublet case,  i.e., 
$T\to \{W^+b, Zt, ht, W^+B\}$, however, as just intimated, we can neglect the $T\to W^+B$ case.  
Therefore,  we study in Figs.~\ref{fig17}--\ref{fig18}, the total SM and non-SM decays of the $T$ state and 
the ${\cal BR}$ correlations already seen, respectively. In the first figure,   we recognize the generally subleading nature of the latter concerning the former. 
In the second figure, with the ${\cal BR}$'s mapped against $\tan\beta$, it is relevant to notice that the triplet patterns change rather drastically concerning the singlet and doublet ones for the case of $T$ decays, showing a strong anti-correlation between standard and exotic channels, for both charged and (especially) neutral currents. As for the $\tan\beta$ dependence, this is such that large values of it favour the SM decays and small values favour 2HDM ones. 	Altogether, the exotic channels can have decay rates ranging from about 15\% to 30\%.

\begin{figure*}[htpb!]
	\centering
	\includegraphics[width=0.45\textwidth,height=0.45\textwidth]{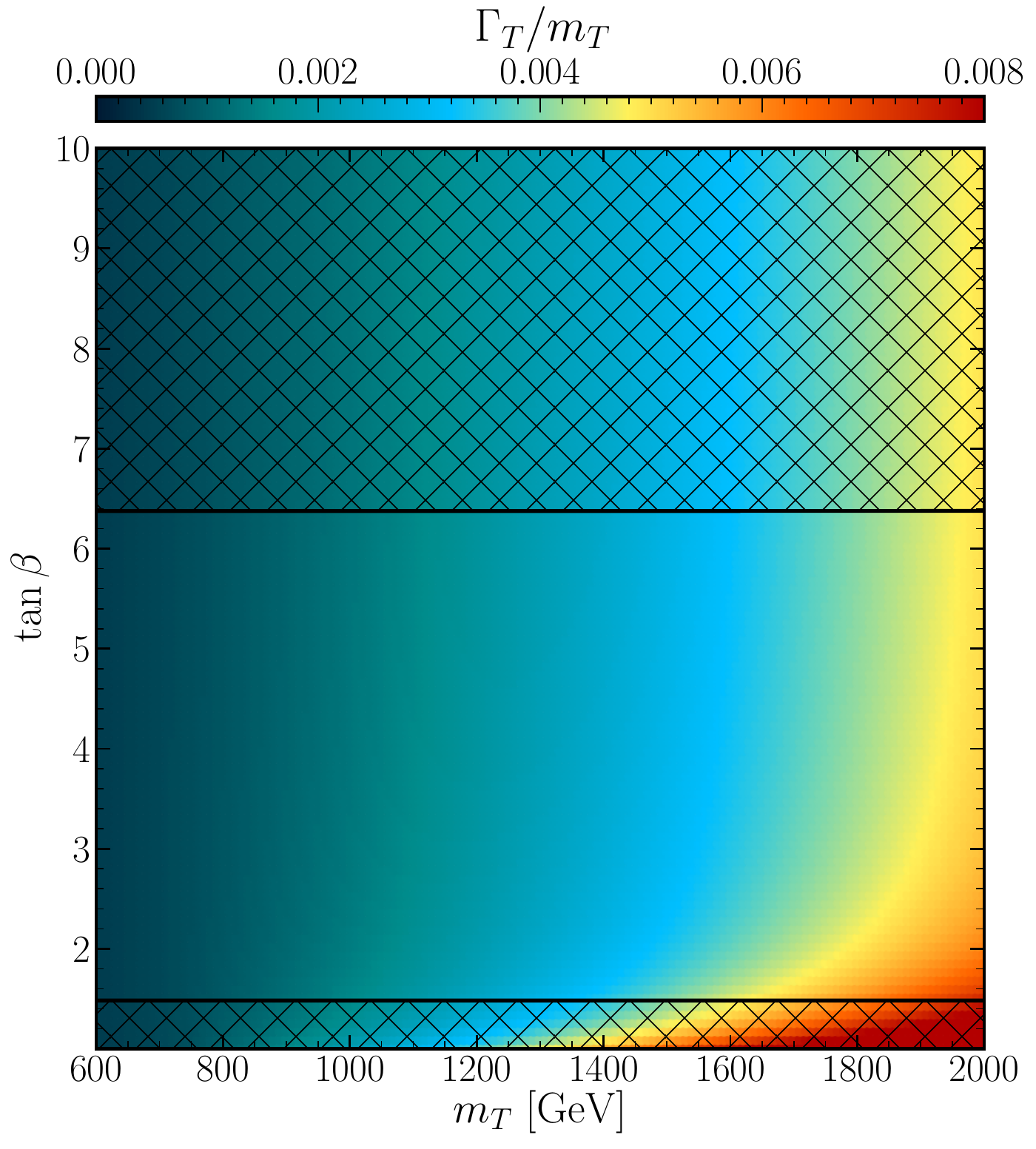} 
	\caption{The ratio $\Gamma_T/m_T$ mapped over the $(m_T,\tan\beta)$ plane.  with $\sin\theta_L^u=0.02$ (the 2HDM parameters are the same as in Fig. \ref{fig3}). Here, the shaded areas are excluded by \texttt{HiggsBounds}, and all other constraints  ($S$, $T$, \texttt{HiggsSignals} and theoretical ones) are also checked.}	
	\label{fig19}
\end{figure*}

Before presenting our BPs, we illustrate in Fig.~\ref{fig19} the ratio $\Gamma_T/m_T$, in order to show that, in this BSM scenario, the $T$ state is always very narrow, in fact, the quantity $\Gamma_T/m_T$ would reach 0.008 at the most and only for high $m_T$. The BPs themselves, detailed in Tab.~\ref{Bp7}, cover a variety of mass spectra for the $T$, enabling a focused investigation into the phenomenology of exotic $T$ decays.

\begin{table*}[htpb!]
	\begin{center}
		\setlength{\tabcolsep}{30pt}
		\renewcommand{\arraystretch}{0.8}
		\begin{adjustbox}{max width=\textwidth}		
			\begin{tabular}{lccc}
				\toprule\toprule
				Parameters &       BP$_1$ &       BP$_2$ &       BP$_3$ \\\toprule
				\multicolumn{4}{c}{2HDM+VLQ inputs. The masses are in GeV.} \\\toprule

				$m_h$   &   125 &    125 &  125\\
				$m_H$  &   687.46 &  745.70 &  717.32\\
				$m_A$   &  573.55 &  743.41 &  701.51\\
				$m_{H\pm}$   &742.23 &  862.22 &  811.92  \\
				$\tan\beta$ &    1.06 &    1.03 &    1.02\\
				$m_T$      & 907.50 & 1611.49 & 1911.18 \\
				$m_B$      &899.36 & 1600.20 & 1895.30 \\
				$m_Y$      & 891.74 & 1589.08 & 1879.56  \\
				$\sin(\theta^u)_L$    & -0.19 &    0.17 &   -0.18  \\
				$\sin(\theta^d)_L$    &-0.13 &    0.12 &   -0.13 \\
				\toprule
				\multicolumn{4}{c}{$\mathcal{BR}(H^\pm\to {XY})$ in \%} \\\toprule
				${\cal BR}(H^+\to t\bar{b})$ &82.66 & 92.82 & 96.46  \\
				${\cal BR}(H^+\to\tau\nu)$ & - &   - &   - \\
				${\cal BR}(H^+\to W^+ A)$ & 17.19 &  3.70 &  2.60 \\
				\toprule
				\multicolumn{4}{c}{$\mathcal{BR}(T\to {XY})$ in \%} \\\toprule
				${\cal BR}(T\to W^+b)$  &  0.11 &  0.01 &  0.00\\
				${\cal BR}(T\to W^+B)$  &   - &   - &   - \\
				${\cal BR}(T\to Zt)$  &  37.98 & 24.40 & 21.15 \\
				${\cal BR}(T\to ht)$  &41.89 & 25.18 & 21.62  \\
				${\cal BR}(T\to Ht)$  & 4.81 & 14.03 & 15.01\\
				${\cal BR}(T\to At)$  &   7.72 & 12.56 & 14.14 \\
				${\cal BR}(T\to H^+b)$ & 7.49 & 23.82 & 28.09\\
				${\cal BR}(T\to H^+B)$ &   - &  - &   -  \\
				
				\toprule
				\multicolumn{4}{c}{Total decay width in GeV.} \\\toprule
				$\Gamma(T)$ &11.40 & 77.33 & 172.09\\\toprule
				
				\multicolumn{4}{c}{Observables} \\\toprule
				$T_{\mathrm{2HDM}}$  &   0.1640 &    0.2464 &    0.1859  \\
				$T_{\mathrm{VLQ}}$  & -0.1366 &   -0.1373 &   -0.1006 \\
				$S_{\mathrm{2HDM}}$ &-0.0083 &   -0.0076 &   -0.0070\\
				$S_{\mathrm{VLQ}}$ &  0.0239 &    0.0200 &    0.0238 \\	
				$\Delta\chi^2(S_{\mathrm{2HDM+VLQ}},T_{\mathrm{2HDM+VLQ}})$ &    1.81 &    3.45 &    0.82 \\\toprule
				$\chi^2{(h_{125})}\equiv\chi^2_\texttt{HiggsSignals}$ & 158.55 &  158.43 &  157.82 \\	
				\toprule\toprule
				
			\end{tabular}
		\end{adjustbox}
	\end{center}
	\caption{The full description of our BPs for the $(TBY)$ triplet case.}\label{Bp7}
\end{table*}

\section{Conclusions}

The top quark plays a key role in SM dynamics, as it is responsible for the instability of the Higgs mass under radiative corrections, for onsetting the so-called hierarchy problem as well
as for the (apparent) metastability of its vacuum. Hence, it is unsurprising that this particle has been heralded as the key messenger of BSM physics. A plausible scenario is that such a state of the SM has one or more companions, i.e., coloured fermionic particles with EM charge $+2/3$, heavier than the top quark itself. However, such new states may not have the same EW properties as the SM object (notably, they cannot be chiral states, i.e., interact with the $W^\pm$ according to a $(1-\gamma_5)$ current), as this would generate problems about the properties of the discovered Higgs state, both in the SM and in extended Higgs models which possess the SM limit. An example of the latter is the 2HDM, wherein VLQs can be added to the extended Higgs spectrum, notably, with the same EM charge as the top quark.  

Specifically, herein, we have considered a 2HDM of Type-II supplemented by a VLQ companion to the top quark ($T$), alongside other new fermionic states ($B$, $X$ and $Y$), of the SM falling in a singlet, doublet or triplet representation. After constraining the parameter space of this model for all VLQ multiplet cases against theoretical and experimental constraints, we have proceeded to study the `standard' decay $T\to W^+b, Zt$ and $ht$ (with $h$ the discovered SM-like Higgs state) as well as the `exotic' ones $T\to H^+b, At$ and $Ht$ (where $H^+,A$ and $H$ are the heavy Higgs states of the 2HDM). $B$ states were also considered amongst the decay products of the $T$ ones, yet the corresponding channels were typically found negligible.

Our conclusions are drawn from a comprehensive analysis conducted at the ${\cal BR}$ level, without any Monte Carlo simulations. Particularly, our exploration of parameter space has revealed promising prospects that open up pathways\\ 
wherein $T$, $H^+$, $A$, and $H$ particles can be light enough to be pursued in direct searches at the LHC, following $T\overline T$ pair production via QCD. Additionally, our investigation into the exotic decay modes of $T$  and the assessment of their relative rates have yielded invaluable insights into the underlying structure of the VLQ multiplet. This enhanced understanding significantly contributes to characterizing the BSM scenario, including confirming its Type-II nature and measuring key parameters within our 2HDM+VLQ model.

Furthermore, we advocate for further exploration through Monte Carlo simulations as the next step in our research program. This initiative aims to rigorously test the feasibility of both an extended Higgs sector and an augmented fermionic spectrum being simultaneously accessible at the LHC through non-SM signatures of VLQs produced in pairs in QCD collisions. To this end, we have proposed several BPs amenable to experimental investigation during Run 3 of the LHC and/or its high luminosity phase.

\section*{Acknowledgments}
We thank R. Enberg, L. Panizzi and S. Taj for their collaboration in the initial stages of this work. The authors have been supported by the grant H2020-MSCA-RISE-2014 no. 645722
(NonMinimalHiggs) in the initial stages of this collaboration.
SM is supported in part through the NExT Institute, the STFC Consolidated Grant ST/L000296/1 and the Knut and Alice Wallenberg foundation under the
grant KAW 2017.0100 (SHIFT).

\renewcommand{\thetable}{\Roman{table}} 
\begin{strip}
\section*{Appendix}
\subsection{Lagrangian in the mass basis}

After EWSB, we are left with five Higgs bosons that are two-CP even $h$ and $H$, one CP-odd $A$, and then a pair of charged Higgs $H^\pm$. We now collect the Lagrangian on a mass basis in the general 2HDM. 
\subsection{Light-light interactions}
\centering

	\begin{eqnarray}
\mathcal{L}_W &=& - \frac{g}{\sqrt{2}} \overline{t} \gamma^\mu (V^L_{tb}P_L + V^R_{tb} P_R)b W^+_\mu + h.c., \nonumber\\
\mathcal{L}_Z &=& - \frac{g}{2c_W}	 \overline{t} \gamma^\mu ( X^L_{tt}P_L + X^R_{tt}P_R - 2 Q_t s_W^2 ) t Z_\mu \nonumber \\ 
&& - \frac{g}{2c_W}	 \overline{b} \gamma^\mu (- X^L_{bb}P_L - X^R_{bb}P_R - 2 Q_b s_W^2 ) b Z_\mu   + h.c., \nonumber\\
\mathcal{L}_{h} &=& - \frac{g m_t}{2 M_W} Y_{tt}^h \overline{t} t h- \frac{g m_b}{2 M_W} Y_{bb}^h \overline{b} b h  + h.c., \nonumber\\
\mathcal{L}_{H} &=& -  \frac{g m_t}{2 M_W} Y_{tt}^H \overline{t} t H- \frac{g m_b}{2 M_W} Y_{bb}^H \overline{b} b H + h.c.,\nonumber\\
\mathcal{L}_{A} &=& - i \frac{g m_t}{2 M_W} Y_{tt}^A \overline{t}  \gamma_5 t A + i\frac{g m_b}{2 M_W} Y_{bb}^A \overline{b} \gamma_5 b A  + h.c., \nonumber \\
\mathcal{L}_{H^+} &=& -\frac{gm_t}{ \sqrt{2} M_W} \overline{t} (\cot\beta Z^L_{tb} P_L + \tan\beta Z^R_{tb}P_R)b H^+ + h.c.
\end{eqnarray}

\begin{eqnarray}
\begin{array}{c|ccc}
& V^L_{tb} && V^R_{tb} \\ \hline
(T)	& c_L &&0  \\ 
(XT)	& c_L && 0 \\ 
(TB)&  c_L^u c_L^d + s_L^u s_L^d e^{i(\phi_u - \phi_d)}   && s_R^u s_R^d e^{i(\phi_u - \phi_d)} \\
(XTB)&  c_L^u c_L^d + \sqrt{2} s_L^u s_L^d   && \sqrt{2} s_R^u s_R^d    \\ 
(TBY)&  c_L^u c_L^d + \sqrt{2} s_L^u s_L^d    && \sqrt{2} s_R^u s_R^d  
\end{array} 
\nonumber
\end{eqnarray}	

\captionof{table}{Light-light couplings to the $W$ boson.}	

\begin{eqnarray}
\begin{array}{c|cccccc}
& \quad	X_{tt}^L &&\quad	X_{tt}^R&\quad	X_{bb}^L  && \quad X_{bb}^R  \\ \hline
(T)	& c_L^2 && 0  &1 && 0\\
(XT)	& c_L^2 - s_L^2 && -s_R^2  &1 && 0\\
(TB) & 1 && (s_R^u)^2  &1 && (s_R^d)^2 \\
(XTB)	& (c_L^u)^2 && 0  & 1 + (s_L^d)^2  && 2 (s_R^d)^2\\
(TBY)	& 1 + (s_L^u)^2   && 2 (s_R^u)^2  & (c_L^d)^2 && 0
\end{array} \nonumber
\end{eqnarray}\captionof{table}{Light-light couplings to the $Z$ boson.}

\begin{eqnarray}
\begin{array}{c|ccc }
&  	Y_{tt}^h & 	Y_{tt}^H& 	Y_{tt}^A    \\ \hline
(T)	& (s_{\beta\alpha}  +    c_{\beta\alpha} \cot\beta) c_L^2 &  (c_{\beta\alpha}  -    s_{\beta\alpha} \cot\beta) c_L^2  & -\cot\beta c_L^2   \\
(XT)	& (s_{\beta\alpha}  +    c_{\beta\alpha} \cot\beta) c_R^2 &  (c_{\beta\alpha}  -    s_{\beta\alpha} \cot\beta) c_R^2  & -\cot\beta c_R^2   \\
(TB) & (s_{\beta\alpha}  +    c_{\beta\alpha} \cot\beta) (c_R^u)^2 & (c_{\beta\alpha}  -    s_{\beta\alpha} \cot\beta) (c_R^u)^2  &-\cot\beta (c_R^u)^2\\
(XTB)	& (s_{\beta\alpha}  +    c_{\beta\alpha} \cot\beta)(c_L^u)^2    &  (c_{\beta\alpha}  -    s_{\beta\alpha} \cot\beta) (c_L^u)^2    & -\cot\beta (c_L^u)^2   \\
(TBY)	& (s_{\beta\alpha}  +    c_{\beta\alpha} \cot\beta)(c_L^u)^2    &  (c_{\beta\alpha}  -    s_{\beta\alpha} \cot\beta) (c_L^u)^2    & -\cot\beta (c_L^u)^2   
\end{array} \nonumber
\end{eqnarray}\captionof{table}{Light-light top quark couplings to the triplets Higgs \{$h,H,A$\}.}

\begin{eqnarray}
\begin{array}{c|ccc }
&  	Y_{bb}^h & 	Y_{bb}^H& 	Y_{bb}^A    \\ \hline
(T)	& s_{\beta\alpha}  -    c_{\beta\alpha} \tan\beta &c_{\beta\alpha}  +   s_{\beta\alpha} \tan\beta & \tan\beta \\
(XT)	& s_{\beta\alpha}  -    c_{\beta\alpha} \tan\beta &c_{\beta\alpha}  +   s_{\beta\alpha} \tan\beta & \tan\beta \\
(TB) & (s_{\beta\alpha}  -    c_{\beta\alpha} \tan\beta) (c_R^d)^2 & (c_{\beta\alpha}  +   s_{\beta\alpha} \tan\beta) (c_R^d)^2  & \tan\beta (c_R^d)^2\\
(XTB)	& (s_{\beta\alpha}  -    c_{\beta\alpha} \tan\beta) (c_L^d)^2 &  (c_{\beta\alpha}  +   s_{\beta\alpha} \tan\beta) (c_L^d)^2  & \tan\beta (c_L^d)^2 \\
(TBY)	& (s_{\beta\alpha}  -    c_{\beta\alpha} \tan\beta) (c_L^d)^2 &  (c_{\beta\alpha}  +   s_{\beta\alpha} \tan\beta) (c_L^d)^2  & \tan\beta (c_L^d)^2 \\
\end{array} \nonumber
\end{eqnarray}\captionof{table}{Light-light bottom quark couplings to the triplets Higgs \{$h,H,A$\}.}

\begin{eqnarray}
\begin{array}{c|ccc}
& Z^L_{tb} && Z^R_{tb} \\ \hline
(T)	&  c_L    &&  \frac{m_b}{m_t}c_L \\ 
(XT)	& c_R   &&  \frac{m_b}{m_t}c_L  \\ 
(TB)&  c_L^d c_L^u + \frac{s_L^d}{s_L^u } (s_L^u{}^2 - s_R^u{}^2) e^{i(\phi_u - \phi_d)}   && \frac{m_b}{m_t}\left[ c_L^u c_L^d + \frac{s_L^u}{s_L^d } (s_L^d{}^2 - s_R^d{}^2) e^{i(\phi_u - \phi_d)} \right]  \\
(XTB)&  c_L^u   &&  \frac{m_b}{m_t}c_L^d   \\ 
(TBY)&     c_L^u   &&  \frac{m_b}{m_t}c_L^d 
\end{array} 
\nonumber
\end{eqnarray}	
\captionof{table}{Light-light couplings to the Higgs charged.}	

\subsection{Heavy-heavy interactions}	

\begin{eqnarray}
\mathcal{L}_W&=& - \frac{g}{\sqrt{2}} \overline{Q} \gamma^\mu (V^L_{QQ^\prime} P_L + V^R_{QQ^\prime} P_R    ) Q^\prime W^+_\mu + h.c.,\nonumber\\
\mathcal{L}_Z &=& - \frac{g}{2 c_W}  \overline{Q} \gamma^\mu (\pm X^L_{QQ}P_L\pm X^R_{QQ}P_R-2Q_Qs_W^2  )Q Z_\mu \nonumber \\\
\mathcal{L}_{h} &=& - \frac{g m_Q}{2 M_W} Y_{QQ}^h \overline{Q} Q h  + h.c., \nonumber\\
\mathcal{L}_{H} &=& -  \frac{g m_Q}{2 M_W} Y_{QQ}^H \overline{Q} Q H  + h.c.,\nonumber\\
\mathcal{L}_{A} &=& \pm i \frac{g m_Q}{2 M_W} Y_{QQ}^A \overline{Q}  \gamma_5 Q A  + h.c., \nonumber \\
\mathcal{L}_{H^+} &=& -\frac{gm_Q}{ \sqrt{2} M_W} \overline{Q} (\cot\beta Z^L_{QQ} P_L + \tan\beta Z^R_{QQ}P_R)Q H^+ + h.c.
\end{eqnarray}

\begin{eqnarray}
\begin{array}{c|cccccc}
& 	V_{ {T}B}^L &&&&&\quad	V_{ {T}B}^R  \\ \hline
(TB)	 & \quad s_L^u s_L^de^{-i(\phi_u - \phi_d)} + c_L^u c_L^d   &&&&& \quad  c_R^u c_R^d  \\
(XTB)	 & s_L^u s_L^d + \sqrt{2} c_L^u c_L^d &&&&& \quad \sqrt{2} c_R^u c_R^d  \\
(TBY)    & s_L^u s_L^d + \sqrt{2} c_L^u c_L^d &&&&& \quad \sqrt{2} c_R^u c_R^d 
\end{array} \nonumber
\end{eqnarray}\captionof{table}{Heavy-heavy couplings to the $W$ boson.}	

\begin{eqnarray}
\begin{array}{c|cccccc}
&  	X_{TT}^L &&&&&	X_{TT}^R  \\ \hline
(T)	& (s_L)^2  &&&&& 0   \\

(XT)& s_L^2 - c_L^2  &&&&& - c_R^2  \\
(TB)&1  &&&&&(c_R^u)^2    \\
(XTB)& (s_L^u)^2  &&&&& 0   \\
(TBY)& 1 + (c_L^u)^2   &&&&&   2 (c_R^u)^2 
\end{array} \nonumber
\end{eqnarray}\captionof{table}{Heavy-heavy couplings to the $Z$ boson.}

\begin{eqnarray}
\begin{array}{c|ccc}
&  	Y_{TT}^h & 	Y_{TT}^H& 	Y_{TT}^A     \\ \hline
(T)	& (s_{\beta\alpha}  +    c_{\beta\alpha} \cot\beta) s_L^2 &  (c_{\beta\alpha}  -    s_{\beta\alpha} \cot\beta) s_L^2  & -\cot\beta s_L^2   \\  
(XT)	& (s_{\beta\alpha}  +    c_{\beta\alpha} \cot\beta) s_R^2 &  (c_{\beta\alpha}  -    s_{\beta\alpha} \cot\beta) s_R^2  & -\cot\beta s_R^2   \\  
(TB) &(s_{\beta\alpha}  +    c_{\beta\alpha} \cot\beta) (s_R^u)^2 &  (c_{\beta\alpha}  -    s_{\beta\alpha} \cot\beta) (s_R^u)^2  & -\cot\beta  (s_R^u)^2 \\  
(XTB) &(s_{\beta\alpha}  +    c_{\beta\alpha} \cot\beta)  (s_L^u)^2 &  (c_{\beta\alpha}  -    s_{\beta\alpha} \cot\beta) (s_L^u)^2  & -\cot\beta  (s_L^u)^2 \\  
(TBY) &(s_{\beta\alpha}  +    c_{\beta\alpha} \cot\beta)  (s_L^u)^2 &  (c_{\beta\alpha}  -    s_{\beta\alpha} \cot\beta) (s_L^u)^2  & -\cot\beta  (s_L^u)^2 
\end{array} \nonumber\label{tabXIV}
\end{eqnarray}\captionof{table}{Heavy-heavy Top VLQ couplings to the triplets Higgs \{$h,H,A$\}.}	

\begin{eqnarray}
\begin{array}{c|ccc}
& Z^L_{TB} && Z^R_{TB} \\ \hline 
(TB)&  s_L^d s_L^u e^{i(\phi_d - \phi_u)}+ \frac{ c_L^d}{c_L^u} (s_R^u{}^2  - s_L^u{}^2 ) &&\frac{m_B}{ m_T}   \left[s_L^u s_L^d e^{i(\phi_d - \phi_u)}+  \frac{c_L^u}{ c_L^d} ( s_R^d{}^2  -  s_L^d{}^2 )  \right]\\ 
(XTB)& -   && -  \\ 
(TBY)&-   &&- 
\end{array} 
\nonumber
\end{eqnarray}	
\captionof{table}{Heavy-heavy couplings to the Higgs charged.}

\subsection{Light-heavy interactions}

\begin{eqnarray}
\mathcal{L}_W & = & - \frac{g}{\sqrt{2}} \overline{Q} \gamma^\mu (V^L_{Qq}P_L + V^R_{Qq}P_R )q W^+_\mu \nonumber\\
&& - \frac{g}{\sqrt{2}} \overline{q} \gamma^\mu (V^L_{qQ}P_L + V^R_{qQ}P_R )Q W^+_\mu   + h.c.\nonumber \\
\mathcal{L}_Z&=& - \frac{g}{2 c_W} \overline{q} \gamma^\mu (\pm X^L_{qQ}P_L \pm X^R_{qQ}P_R )QZ_\mu + H.c  \nonumber \\
\mathcal{L}_{h} & = & - \frac{g m_T}{2 M_W} \overline{t} ( Y^L_{htT} P_L + Y^R_{htT}P_R)Th\nonumber \\
&&- \frac{gm_B}{2M_W} \overline{b} (Y^L_{hbB} P_L+ Y^R_{hbB} P_R ) B h+ H.C.\nonumber\\
\mathcal{L}_{H} & = &   -\frac{g m_T}{2 M_W}  \overline{t} ( Y^L_{HtT} P_L + Y^R_{HtT}P_R)T H\nonumber \\
&&- \frac{gm_B}{2M_W}   \overline{b} (Y^L_{HbB} P_L+ Y^R_{HbB} P_R ) B h+ H.C. \nonumber\\
\mathcal{L}_{A} & = &   i\frac{g m_T}{2 M_W}  \overline{t} ( Y^L_{AtT} P_L - Y^R_{AtT}P_R)T A \nonumber \\
&&- i\frac{gm_B}{2M_W}  \overline{b} (Y^L_{AbB} P_L -  Y^R_{AbB} P_R ) B A + H.C. \nonumber \\
\mathcal{L}_{H^+} &=& - \frac{g m_T}{\sqrt{2}M_W}\overline{T} (\cot\beta Z^L_{Tb} P_L + \tan\beta Z^R_{Tb} P_R ) b H^+\nonumber\\
&&  - \frac{g m_B}{\sqrt{2}M_W}\overline{t} (\cot\beta Z^L_{tB} P_L + \tan\beta Z^R_{tB} P_R ) B H^+ +H.c,
\end{eqnarray}

\begin{eqnarray}
\begin{array}{c|cccccc}
& 	V_{Tb}^L &&&&&	V_{Tb}^R    \\ \hline
(T)	&s_L  e^{-i\phi}   &&&&& 0 \\ 
(XT) & s_L  e^{-i\phi}  &&&&& 0 \\
(TB) &s_L^u c_L^d e^{-i\phi_u}- c_L^u s_L^d e^{-i\phi_d}  &&&&& - c_R^u s_R^d e^{-i\phi_d}\\
(XTB) & (s_L^u c_L^d - \sqrt{2} c_L^u s_L^d )  e^{-i\phi}  &&&&& - \sqrt{2} c_R^u s_R^d e^{-i\phi}\\
(TBY) & (s_L^u c_L^d - \sqrt{2} c_L^u s_L^d )  e^{-i\phi}  &&&&& - \sqrt{2} c_R^u s_R^d e^{-i\phi}
\end{array} \nonumber 
\end{eqnarray}\captionof{table}{Heavy-light couplings to the $W$ boson.}	

\begin{eqnarray}
\begin{array}{c|cccccc}
& 	X_{tT}^L &&&&&	X_{tT}^R  \\ \hline
(T)	&  c_L s_L e^{i\phi} &&&&& 0  \\
(XT)	& 2 c_L s_L e^{i\phi} &&&&&  c_R s_R e^{i\phi}  \\
(TB)	&0 &&&&& 	-s_R^u c_R^u e^{i\phi_u} \\
(XTB)	& s_L^u c_L^u e^{i\phi} &&&&& 0 \\
(TBY)	& -s_L^u c_L^u e^{i\phi} &&&&&  -2s_R^u c_R^u e^{i\phi} 
\end{array} \nonumber\label{t2}
\end{eqnarray}\captionof{table}{Light-heavy couplings to the $Z$ boson.}	 
\begin{eqnarray}
\begin{array}{c|ccc }
&  	Y_{htT}^L & 	Y_{HtT}^L& 	Y_{AtT}^L    \\ \hline
(T)	& (s_{\beta\alpha}  +    c_{\beta\alpha} \cot\beta)\frac{m_t}{m_T} c_L s_L e^{i\phi} &  (c_{\beta\alpha}  -    s_{\beta\alpha} \cot\beta)\frac{m_t}{m_T} c_L s_L e^{i\phi}   & -\cot\beta\frac{m_t}{m_T} c_L s_L e^{i\phi}   \\
(XT)	& (s_{\beta\alpha}  +    c_{\beta\alpha} \cot\beta) c_R s_R e^{i\phi} &  (c_{\beta\alpha}  -    s_{\beta\alpha} \cot\beta)  c_R s_R e^{i\phi}   & -\cot\beta  c_R s_R e^{i\phi}   \\ 
(TB) &  (s_{\beta\alpha}  +    c_{\beta\alpha} \cot\beta) s_R^u c_R^u e^{i\phi_u} &  (c_{\beta\alpha}  -    s_{\beta\alpha} \cot\beta)  s_R^u c_R^u e^{i\phi_u}  &  -\cot\beta s_R^u c_R^u e^{i\phi_u}\\
(XTB)	& (s_{\beta\alpha}  +    c_{\beta\alpha} \cot\beta)\frac{m_t}{m_T} s_L^u c_L^u e^{i\phi} &  (c_{\beta\alpha}  -    s_{\beta\alpha} \cot\beta)\frac{m_t}{m_T} s_L^u c_L^u e^{i\phi}   & -\cot\beta\frac{m_t}{m_T} s_L^u c_L^u e^{i\phi}   \\
(TBY)	& (s_{\beta\alpha}  +    c_{\beta\alpha} \cot\beta)\frac{m_t}{m_T} s_L^u c_L^u e^{i\phi} &  (c_{\beta\alpha}  -    s_{\beta\alpha} \cot\beta)\frac{m_t}{m_T} s_L^u c_L^u e^{i\phi}   & -\cot\beta\frac{m_t}{m_T}s_L^u c_L^u e^{i\phi}   \\
\end{array} \nonumber\label{XVIII}
\end{eqnarray}\captionof{table}{Light-heavy left couplings of Top quarks to the triplets Higgs \{$h,H,A$\}.}	

\begin{eqnarray}
\begin{array}{c|ccc }
&  	Y_{htT}^R & 	Y_{HtT}^R & 	Y_{AtT}^R   \\ \hline
(T)	& (s_{\beta\alpha}  +    c_{\beta\alpha} \cot\beta) c_L s_L e^{i\phi} &  (c_{\beta\alpha}  -    s_{\beta\alpha} \cot\beta) c_L s_L e^{i\phi}   & -\cot\beta c_L s_L e^{i\phi}   \\ 
(XT)	& (s_{\beta\alpha}  +    c_{\beta\alpha} \cot\beta)\frac{m_t}{m_T} c_R s_R e^{i\phi} &  (c_{\beta\alpha}  -    s_{\beta\alpha} \cot\beta)\frac{m_t}{m_T}c_R s_R e^{i\phi}   & -\cot\beta\frac{m_t}{m_T} c_R s_R e^{i\phi}   \\
(TB) & (s_{\beta\alpha}  +    c_{\beta\alpha} \cot\beta) \frac{m_t}{m_T} s_R^u c_R^u e^{i\phi_u} & (c_{\beta\alpha}  -    s_{\beta\alpha} \cot\beta) \frac{m_t}{m_T} s_R^u c_R^u e^{i\phi_u}  & -\cot\beta \frac{m_t}{m_T} s_R^u c_R^u e^{i\phi_u}\\
(XTB)	& (s_{\beta\alpha}  +    c_{\beta\alpha} \cot\beta) s_L^u c_L^u e^{i\phi} &  (c_{\beta\alpha}  -    s_{\beta\alpha} \cot\beta) s_L^u c_L^u e^{i\phi}   & -\cot\beta s_L^u c_L^u e^{i\phi}   \\ 
(TBY)	& (s_{\beta\alpha}  +    c_{\beta\alpha} \cot\beta) s_L^u c_L^u e^{i\phi} &  (c_{\beta\alpha}  -    s_{\beta\alpha} \cot\beta) s_L^u c_L^u e^{i\phi}   & -\cot\beta s_L^u c_L^u e^{i\phi}   \\ 
\end{array} \nonumber\label{XIX}
\end{eqnarray}\captionof{table}{Light-heavy right couplings of Top quarks to the triplets Higgs \{$h,H,A$\}.}

\begin{eqnarray}
\begin{array}{c|ccc}
& Z^L_{Tb} && Z^R_{Tb} \\ \hline 
(T)& s_L &&  \frac{m_b}{m_T}s_L \\ 
(XT)&0&&0    \\ 
(TB)&  c_L^d s_L^u e^{- i\phi_u}  +  ( s_L^u{}^2 -  s_R^u{}^2  )\frac{s_L^d}{c_L^u}  e^{- i\phi_d}  && \frac{m_b}{ m_T} \left[    c_L^d s_L^u e^{ -i\phi_u}+  (s_R^d{}^2   -  s_L^d{}^2 ) \frac{c_L^u}{s_L^d}  e^{-i\phi_d}    \right]    \\
(XTB) &s_L^u&&0 
\end{array}  
\nonumber\label{XI}
\end{eqnarray}	
\captionof{table}{Heavy-light couplings to the Higgs charged.}
\end{strip}

\section*{Projections of Electroweak Precision Constraints}

\begin{figure*}[h!]
	\centering
	\includegraphics[width=0.9\textwidth,height=1.3\textwidth]{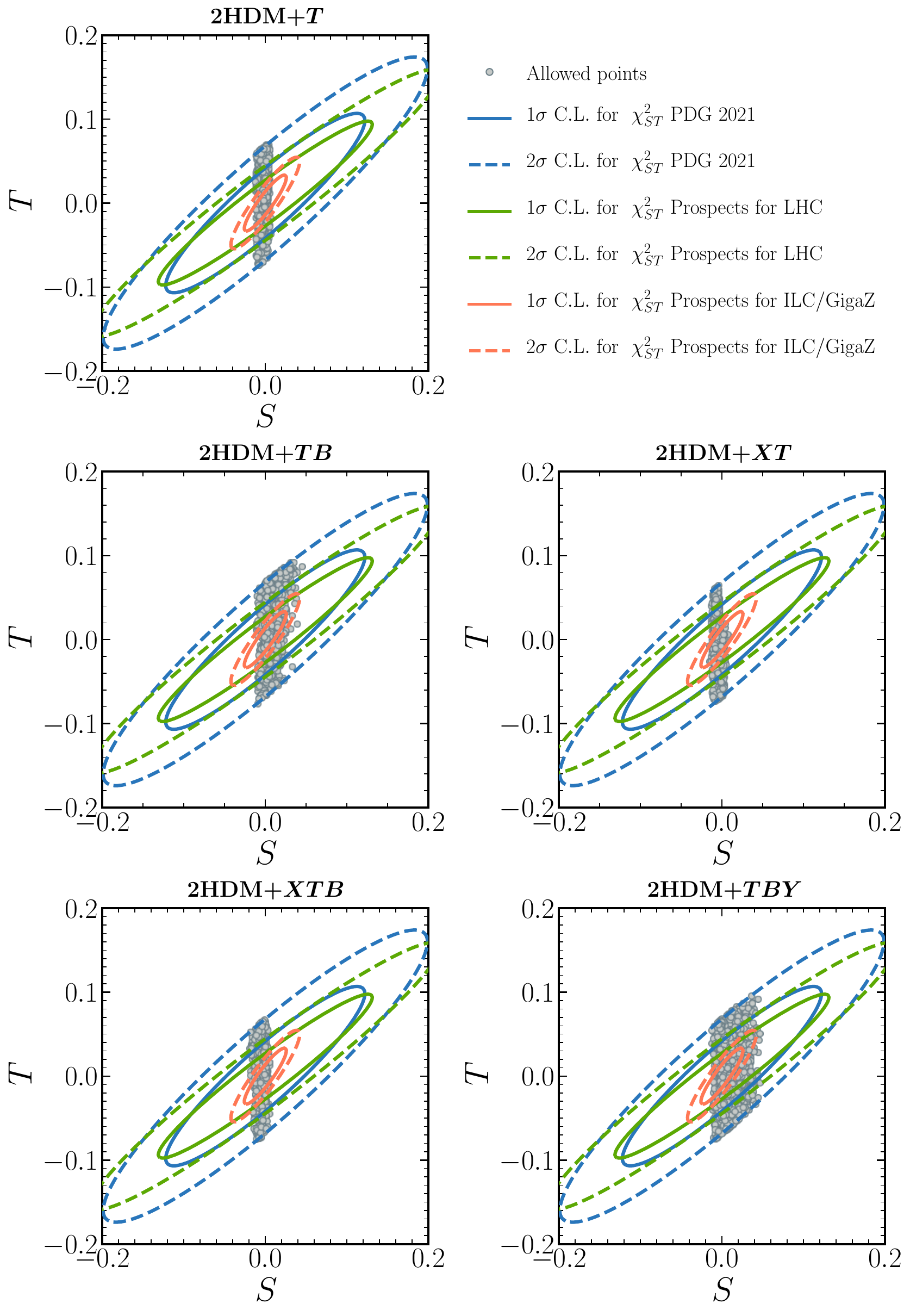}
	\caption{Projections for the electroweak oblique parameters $(S, T)$ for each VLQ representation. The green ellipses indicate the anticipated $1\sigma$ (solid) and $2\sigma$ (dashed) sensitivity regions for future LHC measurements, while the orange ellipses represent the expected sensitivities for the ILC/GigaZ, as provided in \cite{Baak:2014ora}. Both sets of projections assume $U = 0$ and are centred at $S = T = 0$. All points shown satisfy $\chi^2_{(S, T)}(\text{PDG}) \leq 6.18$.}
	\label{fig:ST}
\end{figure*}

In this section, we present the projected electroweak precision constraints on our framework in the $(S, T)$ plane for all VLQ representations, depicted in Fig.~\ref{fig:ST}. This figure illustrates the compatibility of our model’s parameter space with the enhanced precision expected from future measurements at the LHC and ILC/GigaZ \cite{Baak:2014ora}. The green ellipses denote the anticipated $1\sigma$ (solid) and $2\sigma$ (dashed) confidence regions from prospective LHC data, while the orange ellipses represent the projected sensitivities from the ILC/GigaZ. Both projections are derived under the assumption that $U = 0$ and are centred at $S = T = 0$, aligned with the SM reference point.

As illustrated, all points with $\chi^2_{(S, T)}(\text{PDG}) \leq 6.18$ lie within the projected bounds, confirming that our parameter space is robust against future electroweak precision constraints anticipated from upcoming collider experiments.

\bibliographystyle{JHEP}
\bibliography{main}

\end{document}